\documentclass{aa}  

\usepackage{graphicx}
\usepackage{txfonts}
\usepackage{lipsum}
\usepackage{subcaption}
\usepackage{hyperref}
\usepackage{natbib}
\usepackage{tikz}
\usepackage{xcolor}
\usepackage{threeparttable}
\usepackage{comment}
\usetikzlibrary{shapes.geometric, arrows, positioning}

\definecolor{darkyellow}{RGB}{254,153,0}

\tikzstyle{box} = [rectangle, rounded corners, minimum width=5cm, minimum height=1.5cm, text centered, draw=black, fill=yellow!60, text=black]
\tikzstyle{smallbox} = [rectangle, rounded corners, minimum width=5cm, minimum height=1cm, text centered, draw=black, fill=yellow!40, text=black]
\tikzstyle{arrow} = [thick,->,>=stealth, black]

\usepackage{lscape}            
\usepackage{placeins}

\begin{document}

   \title{Interpreting the scattering surface in protoplanetary disks}

\author{Massimiliano Bolchini\inst{1,}\inst{2,}\inst{3}
        \and Giovanni Rosotti \inst{1} \and Marion Villenave \inst{5} \and Antonio Garufi \inst{4} \and Myriam Benisty \inst{6} \and Tilman Birnstiel \inst{7,}\inst{8} \and Stefano Facchini \inst{1} \and Leonardo Testi \inst{2}
        }

   \institute{Dipartimento di Fisica, Università degli Studi di Milano, via G. Celoria 16, 20133 Milano, Italy
            \and Department of Physics and Astronomy, University of Bologna, Via Gobetti 93/2, 40129 Bologna, Italy
            \and INAF, Astrophysics and Space Science Observatory Bologna, Via Gobetti 93/3, 40129 Bologna, Italy\\
             \email{massimilian.bolchin2@unibo.it}
             \and Istituto di Radioastronomia, INAF, Bologna, Italy
            \and Univ. Grenoble Alpes, CNRS, IPAG, 38000 Grenoble, France  
            \and Max-Planck-Institut für Astronomie (MPIA), Königstuhl 17, 69117 Heidelberg, Germany
            \and University Observatory, Faculty of Physics, Ludwig-Maximilians-Universität München, Scheinerstr. 1, 81679 Munich, Germany
            \and Exzellenzcluster ORIGINS, Boltzmannstr. 2, 85748 Garching, Germany}
   \date{Received September 30, 20XX}

  \abstract
   {In recent years, extreme adaptive optics have enabled high-resolution, high-contrast scattered-light observations of protoplanetary disks. Interpreting these observations requires an understanding of the scattering surface, which is shaped by the distribution of small dust grains and determines how the disks appear in scattered light.}
   {We aim to exploit measurements of the scattering surface height to directly constrain the masses of small dust grains in disks.} 
   {Starting from radiative transfer principles, we developed a semi-analytical model of the stellar radiation path and its interaction with the disk, deriving the height of the scattering surface as a function of disk parameters, such as the mass, temperature, and opacity. We validated our predictions against the radiative transfer code \texttt{MCFOST}. Using measured scattering heights, we inferred the mass of the dust in small grains and the particle size distribution for a sample of ten disks.}
   {We confirmed prior results indicating that the scattering surface coincides with the surface where the integrated optical depth along the stellar path is on the order of unity. The thermal structure of the disk significantly affects the surface height, while dust settling and anisotropic scattering have minor effects. Applying our model to observations, we measured small dust mass fractions on the order of $10^{-3}$ globally. Using models of the dust opacity, we show this is typical for modest amounts of grain growth ($a_{\text{max}} \gtrsim 0.1mm$) and power-law indices of the grain size distribution $\sim 3-3.5$, as commonly found in grain growth models.} {Scattering height measurements, together with the disk’s thermal structure, help set constraints on the small dust content of protoplanetary disks.}


   \maketitle
\nolinenumbers
\section{Introduction}
In recent years, the advent of extreme adaptive optics (AO) has made it possible to observe disks at high resolution in visible and near infrared (NIR) light \citep{Benisty_2023}. In this wavelength range, we see the stellar light that scatters off the small submicron to micron (sub-$\mu$m to $\mu$m) sized grains, whose scattering opacity dominates \citep{Hashimoto_2011, Garufi_2014} over the large grains. These small grains are fundamental in shaping the disk evolution, as they control how stellar radiation is absorbed and re-emitted, shaping the thermal structure of the disk \citep{Chiang_Goldreich_1997, Dullemond_2001, Paola_Dalessio_1999} and its chemistry \citep{Oberg_2023}. Their abundance and size distribution therefore offer key insights into dust growth, fragmentation, and the early stages of planet formation \citep{Drazkowska_2023}.

 Large surveys have been conducted and are still ongoing, such as SEEDS \citep{Tamura_2016}, DARTTS-S \citep{Avenhaus2018}, Gemini-LIGHTS \citep{Rich_2022}, and SPHERE-DESTINYS \citep{Ginski_2021}, meaning that a large sample of observations is becoming available (more than 100 disks observed \citep{Garufi_2026}. In these observations, disks exhibit a flared surface. The height of the surface can be measured with geometrical methods \citep{Avenhaus2018, deBoer_2016, Ginski_2016} or with detailed radiative transfer modeling \citep{Villenave2019}. When rings are present and sufficiently symmetric  their projected shapes on the sky can be fitted with ellipses to geometrically estimate the scattering surface height.

 While these observations are becoming routine, understanding the physics that sets the scattering surface becomes important for interpreting the structure and dust content of disks. A natural approach is to use numerical radiative transfer models, as done in the literature \citep{Pohl_2017, Villenave2019, Rich_2022, Muro-Arena_2018, Franceschi_2023}. However, given the broad parameter space, full numerical radiative transfer models are computationally expensive and feasible only for individual disks. On the other hand, semi-analytic models would allow us to capture the key physical processes while remaining computationally efficient.  

 Protoplanetary disks dust masses are usually estimated through the mm-emission, in the optically thin approximation \citep[e.g.,][]{Andrews_2018}. While this approach is useful to constrain the bulk dust mass, it is insensitive to the population of small $\mu$m-sized grains that dominate the disk surface layers and as a result, the mass in small grains is poorly constrained, despite the important role they play in setting the disk thermal structure, in controlling the reaction rates and tracing the early stages of grain growth. 
In this work, we link the scattering height with the physical properties of the disk, in particular the small dust mass, with a simple semi-analytic model. Starting from radiative transfer considerations, we develop and validate a model that successfully reproduces the emission surface of mid-inclination protoplanetary disks observed in scattered light. We then invert the model and apply it on scattering height measurements from the literature, deriving the mass in small dust for a sample of ten disks. Section \ref{sec: methods} develops the theoretical model, Sect. \ref{sec: mcfost} validates our model against the numerical radiative transfer code \texttt{MCFOST}. Section \ref{sec: results} describes the inversion and mass estimates. Section \ref{sec: discussion} discusses the implications for disk chemistry and grain evolution, while Sect. \ref{sec: conclusions} summarizes our findings.

\section{Methods}
\label{sec: methods}
In this section, we investigate the basic physics governing the location of the scattering surface, namely, the layer from which most of the emission in scattered light comes. The general expectation that the scattering surface lies close to the $\tau_\ast \approx 1$ surface with respect to stellar photons, where $\tau_*$ is the optical depth from the star to the scattering point (e.g., \citealt{Dullemond_2001}) is well established. However, we aim to identify the key disk parameters that influence the location of the surface and to determine relations between these parameters and the surface height. In Sect. \ref{sec: model} we build a semi-analytical model under extremely simplifying assumptions, and in Sect. \ref{sec: approximations} we test the possible approximations and refine the model.

\subsection{Semi-analytical model}
\label{sec: model}
We considered a thermally stratified disk in hydrostatic equilibrium with inclination $i$ with respect to the observer’s line of sight. The vertical density profile is given by (e.g., \citealt{Rosotti_2020_b,Martire_2024})\begin{equation}
    \rho(z)=\rho_0\frac{c_{s,\text{mid-plane}}^2}{c_s^2(z)}\exp\left(-\int_0^z\frac{\Omega_k^2(z')z' \ dz'}{c_s^2(z)}\right),
\end{equation}
where $\rho_0$ is the midplane density, $\Omega_k=GM_*/(r^2+z^2)^{3/2}$ is the Keplerian frequency, and the sound speed $c_s(z)$ is 
\begin{equation}
    c_s(z)=\sqrt{\frac{k_BT(z)}{\mu m_H}}.
\end{equation}
The disk is a passive irradiated disk, for which the temperature structure is set by stellar irradiation rather than viscous heating. In this regime, which is appropriate for Class II protoplanetary disks at radii $\geq 1 \hbox{ AU}$ \citep{Dalessio_1998}, the upper layers intercept more stellar radiation than the midplane, resulting in $T$ increasing with height and a cold midplane \citep{Chiang_Goldreich_1997}.  In the radial direction, $T$ decreases with distance from the star. This vertical and radial temperature structure is directly observed through optically thick CO line emission \citep{Pinte_2018, Law_2021, Law_2022, Law_2023, exoALMA_temperature}. Disks typically show a smooth radial dependence of CO temperature, and following \citet{exoALMA_temperature} and \citet{Law_2021}, the temperature is parametrized as
\begin{equation}
T(z) = 
    \begin{cases}
    T_{atm}+(T_{mid}-T_{atm})\cos^2\Big(\frac{\pi}{2}\frac{z}{z_q}\Big) \ \ \text{for } \ z<z_q \\
    T_{atm } \ \ \ \text{for } \ z<z_q,
    \end{cases}
\end{equation}
where 
\begin{equation}
    T_{atm}(r)=T_{atm,0} \ \Big(\frac{r}{r_0}\Big)^{q_{atm}},
\end{equation}
\begin{equation}
    T_{mid}(r)=T_{mid,0} \ \Big(\frac{r}{r_0}\Big)^{q_{mid}},
\end{equation}
and 
\begin{equation}
    z_q(r)=z_0 \ \Big(\frac{r}{r_0}\Big)^{\beta}.
\end{equation}
Here, we are dealing with typical temperatures on the order of $30$ to $60$ K in the atmosphere, and $10$ to $30$ K in the midplane. While our model is general, for the examples of this paper we adopt as a reference the vertical temperature profile of the LkCa15 disk measured by \citet{exoALMA_temperature}. This choice serves as a representative example and all results hold for disks with different thermal structures. For the surface density profile we adopt the \citet{LyndenBell_Pringle_1974} profile,
\begin{equation}
    \Sigma(r)=\Sigma_0\left(\frac{r_c}{r}\right)\exp\left(-\frac{r}{r_c}\right),
    \label{eq: surface_density_profile}
\end{equation}
where $\Sigma_0$ is the normalization and $r_c$ is the cutoff radius. The consequences of this assumption are discussed in Appendix \ref{appendix: surface density}, where we show that assuming a different slope of the surface density has only a marginal effect on the results. We chose a range of total disk masses $0.001 \ M_{\odot}\leq M \leq 0.1 \ M_{\odot}$. 

Scattered light observations are conducted in \textit{J} ($1.2 \ \mu\hbox{m}$), \textit{H} ($1.6 \ \mu\hbox{m}$) and \textit{K} ($2.2 \ \mu\hbox{m}$) bands: at these wavelengths the opacity is dominated by grains with size up to $\approx 1 \ \mu$m. To confirm this we consider a grain size distribution $n(a)\propto a^{-q}$ with $q = 3.5$ and $a_{\text{max}}= 1 \hbox{ mm}$; and we compute the grain size $a_{\text{90}}$ that contributes for the 90\% of the opacity by solving 
\begin{equation}
    \frac{\int_{a_{\text{min}}}^{a_{\text{90}}}\kappa(a)a^3n(a)da}{\int_{a_{\text{min}}}^{a_{\text{max}}}\kappa(a)a^3n(a)da} = 0.90,
\end{equation} 
where $n(a)$ is the grain size distribution, $\kappa(a)$ is the opacity as a function of grain size, and the factor $a^3$ weights each size bin by its mass contribution (since the opacity $\kappa$ is defined per unit mass of dust).
We compute $\kappa(a)$ with \texttt{Optool} \citep{Dominik_2021}, using the composition from \citet{Ossenkopf_Henning_1994}, and find $a_{90} \simeq 1$~$\mu$m at a wavelength $\lambda = 1.2$~$\mu$m. Henceforth, we refer to grains with sizes up to $a_{\rm max} = 1$~$\mu$m and a grain size distribution exponent of $q = 3.5$ as small dust. We take the opacity per gram of small dust to be $\kappa = 10^4 \ \text{cm}^2\text{g}^{-1}$ at $\lambda = 1.2 \ \mu$m and we assume the dust to gas ratio, $f_{\text{gas to dust }\leq 1\mu\text{m}}$, to be $10^{-3}-10^{-5}$.  Since we are considering micron-sized grains, we neglected dust settling; this approximation is tested in Sect. \ref{sec:settling}.

\begin{figure}[h!]
    \centering
\includegraphics[width=\hsize]{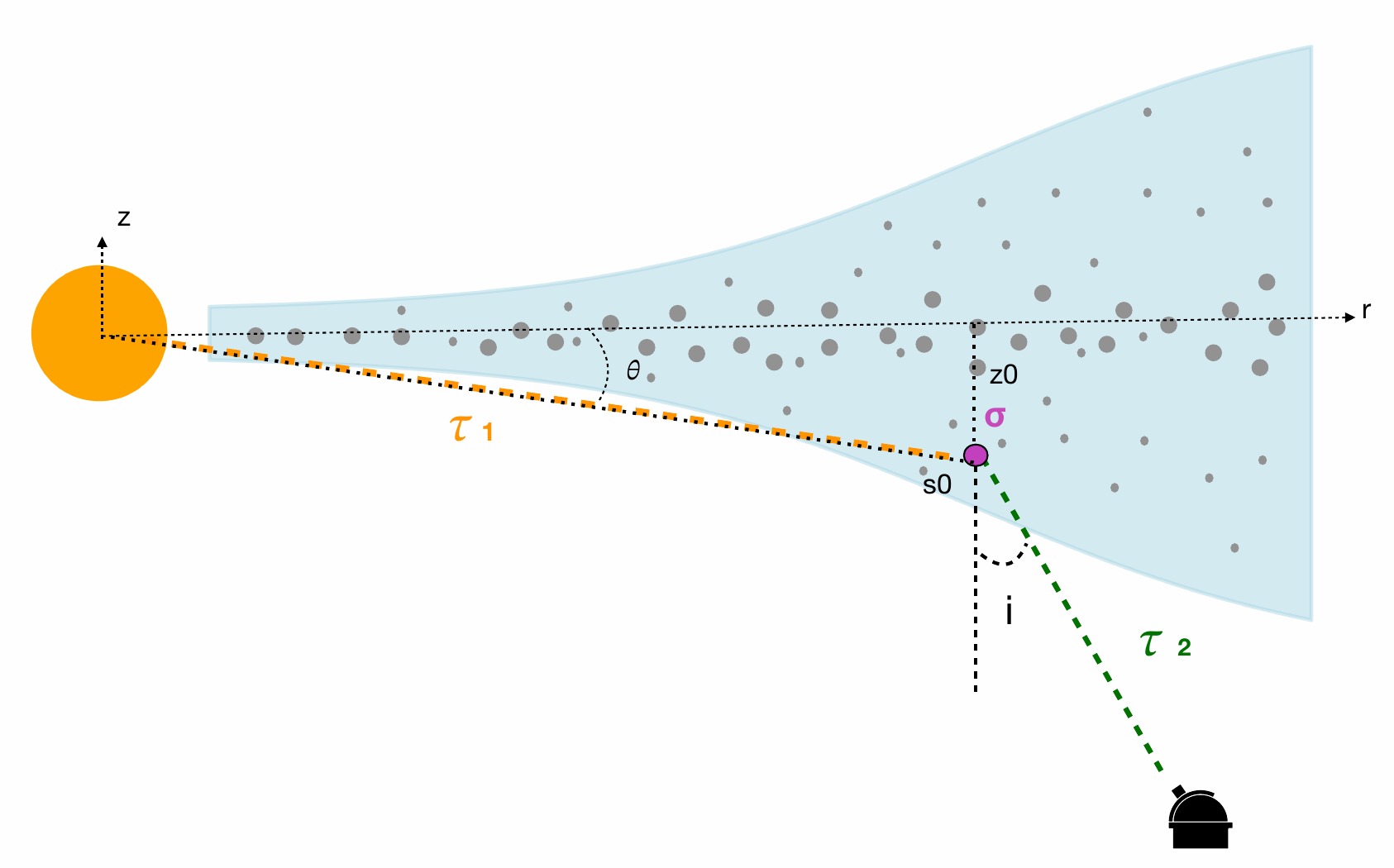}
    \caption{Model of the path of a ray through the disk.}
    \label{fig:model}
\end{figure}

\begin{figure}[h!]
    \centering
\includegraphics[width=\hsize]{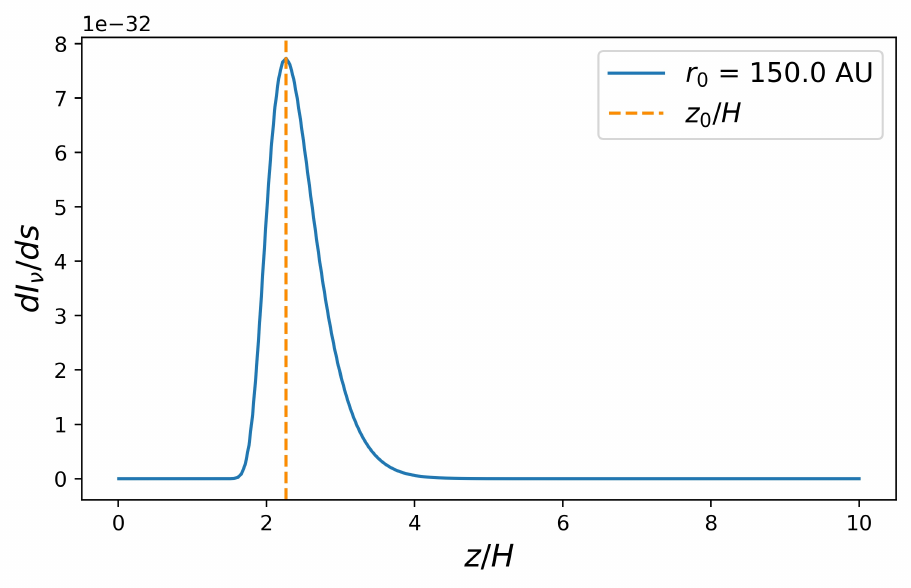}
    \caption{Differential intensity as a function of $z/H$ for $r=150$ AU.}
    \label{fig:dIds}
\end{figure}

To model the stellar radiation propagation through the disk, we adopted the approach of \citet{Paola_Dalessio_1999}, with the single isotropic scattering approximation (we test this approximation in Sect. \ref{sec: mcfost}). In this framework, the intensity along a given ray is given integrating the radiative transfer equation,
\begin{equation}
    \frac{dI_{\nu}}{ds}=\sigma_{\nu}B_{\nu}(T_*)W(r)e^{-\tau_{\nu \ 1}}e^{-\tau_{\nu \ 2}(s)},
    \label{eq: Paola_Dalessio}
\end{equation}
where $s$ is the coordinate along the ray, $\sigma_{\nu} \ [\text{cm}^{-1}]$ is the scattering coefficient, $B_{\nu}$ is the Planck function evaluated at the temperature of the star, and $T_*$, $W(r)=(R_*/r)^2$, is the geometric dilution factor of stellar radiation in the point-source approximation. Then, $\tau_{\nu \  1}$ is the optical depth from the star to the scattering point and $\tau_{\nu \ 2}(s)$ is the optical depth from the scattering point to the observer. As shown in Fig. \ref{fig:model}, we can divide the path into three contributions to the differential intensity:
\begin{itemize}
    \item the path from the star to the scattering point $(r_0,z_0)$, characterized by the optical depth $\tau_1=\int_0^{s_0}\kappa_{\nu}\rho(s)ds$ ($\tau_*$ in \citet{Dullemond_2001});
    \item the scattering at $(r_0,z_0)$, characterized by the cross-section, $\sigma_{\nu}(r_0,z_0)=\kappa_{\nu}\rho(r_0,z_0)$;
    \item the path from the scattering point to the observer at infinity, characterized by the optical depth $\tau_2 = \int_{s_0}^{\infty}\kappa_{\nu}\rho(s)ds$.
\end{itemize}
We are not concerned with the absolute magnitude of the intensity and, instead, we want to identify whether $dI_{\nu}/ds$ as a function of height in the disk has a maximum, which identifies the scattering surface. Specifically, computing $dI_{\nu}/ds$ numerically as a function of the height in the disk $z$, we find a clear maximum (see Fig. \ref{fig:dIds}) and we identify the height $z_0$ of the scattering surface above the midplane as a function of radius.

\begin{figure*}[h!]
    \centering
    \includegraphics[width=\hsize]{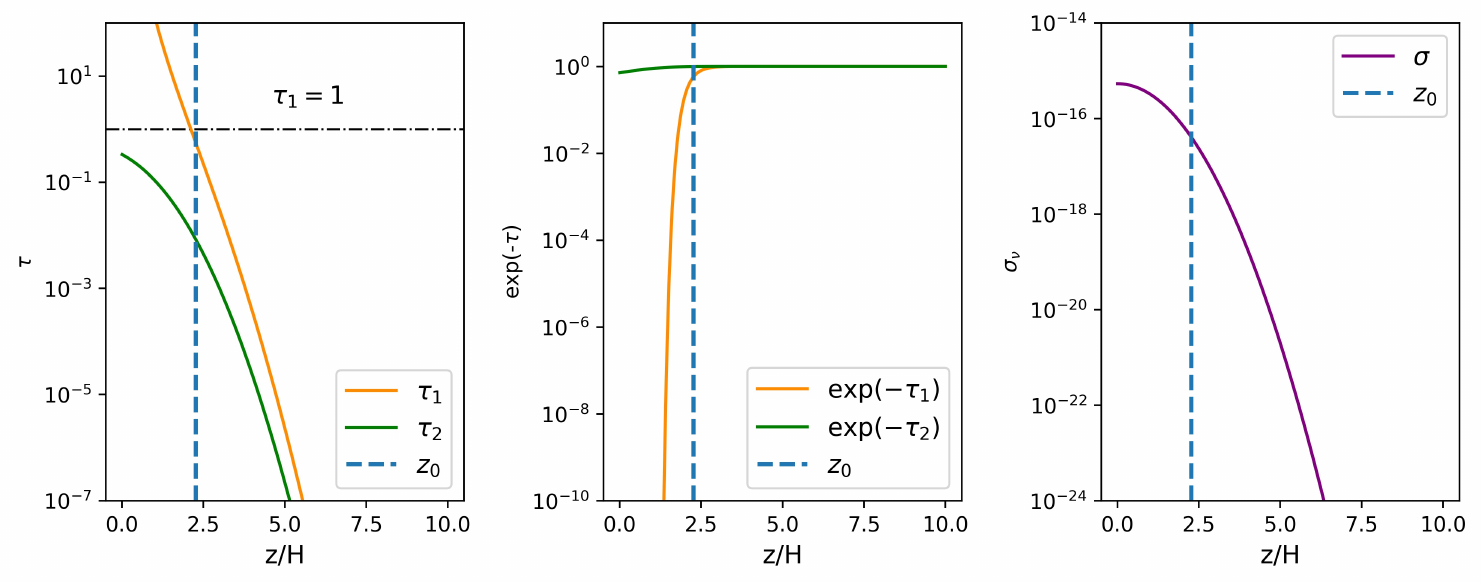}
    \caption{Contributions to the differential intensity as a function of $z$ in the disk at fixed radius $r=150$ AU. \textit{Left}: Optical depth $\tau_1$ and $\tau_2$ as a function of $z/H$. \textit{Middle}: Attenuation factors $e^{-\tau_1}$ and $e^{-\tau_2}$ as a function of $z/H$. \textit{Right}: Scattering cross-section, $\sigma_{\nu}$, as a function of $z/H$.}
    \label{fig:tau_12_sigma}
\end{figure*}
Next, we address what sets the scattering surface. At a fixed radius, it is possible to analyze how different contributions to the optical path vary with height, as illustrated in Fig. \ref{fig:tau_12_sigma}, which shows how these three contributions vary with height in the disk at a fixed radius of 150 AU. The left panel displays the optical depths, $\tau_1$ and $\tau_2$, the central panel shows the corresponding attenuation factors, $e^{-\tau_1}$ and $e^{-\tau_2}$, and the right panel presents the scattering cross-section, $\sigma_{\nu}$. When plotting the scattering height, $z_0$, it becomes evident that this height coincides with the location where $\tau_1\approx 1$ (see left panel). To understand this result, we highlight that there is a competition in setting the scattering surface between the cross-section, $\sigma_{\nu}$ (right panel) and the transmission factor, $e^{-\tau_1}$ (middle panel). At heights lower than the scattering surface, $\tau_1>1$ and the intensity of the stellar radiation is highly attenuated by the  exponential decay, implying that there is little radiation to scatter towards the observer. At heights higher than the scattering surface, the transmission factor flattens out,  and the reduced cross-section implies that although these locations are highly irradiated by the star, most of this radiation simply passes through, with the same net result that little radiation reaches the observer. Consequently, the scattering height is where $\tau_1\approx 1$, maximizing the cross-section, while maintaining sufficient transmission. We note that in general $\tau_2 \ll \tau_1$ and at the scattering surface $\tau_2 \ll 1$; therefore, its effect on the total intensity from the scattering surface is negligible.

To further confirm this result, we computed the scattering height varying the disk mass (ranging from $0.001 M_{\odot}$ to $0.1 M_{\odot}$) and the inclination. We find that this result holds for all the disk masses and for inclinations, $i\leq 60^\circ$. For higher inclinations, $\tau_2$ becomes closer to one and obscures the scattering surface. 

The scattering height can therefore be computed solving the implicit equation, $\tau_1 = 1$:
\begin{multline}
    \tau_1 = \kappa_{\nu}\int\rho(r,z)ds =  \kappa_{\nu}\int_0^{s_0}\frac{\Sigma(r)}{\sqrt{2\pi}H(r)}\cdot \\ \frac{c_{s,\text{mid-plane}}^2}{c_s^2(z)}\exp\left(-\int_0^z\frac{\Omega_k^2(z')z' \ dz'}{c_s^2(z)}\right)ds=1 ,
\end{multline}
where $s$ is the coordinate along the path of the ray, such that $r = s\cos\theta$ and $z = s\sin\theta$, with $\theta=\arctan(z_0/r_0)$. The midplane scale height $H(r)$ is computed taking into account the midplane temperature via 
\begin{equation}
    H(r)=\sqrt{\frac{k_BT_{\text{mid}}(r)}{\mu m_HGM_*}r^3}.
    \label{eq: H_scale}
\end{equation}
The scattering height is therefore determined solving the following equation for $s_0$, namely,
\begin{multline}
    \Sigma_0r_c\sqrt{\frac{\mu m_H G M_*}{k_B}}\int_0^{s_0}\frac{\sqrt{T_{\text{mid}}(r)}}{T(r,z)r^{5/2}}\exp\left(-\frac{r}{r_c}\right)\cdot \\ 
    \cdot\exp\left[-\frac{\mu m_HGM_*}{k_b}\int_0^z\frac{z' \ dz'}{(r^2+z'^2)^{3/2}T(r,z')}\right]ds =1.
    \label{eq: semi-analytical model}
\end{multline}

Finally, it is important to note that the determination of the scattering surface is a nonlocal problem. The height of the surface at a given radius depends not only on the local column density but also on the integrated density along the entire optical path from the star to the scattering point. This highlights the role of the global disk structure in shaping the observed scattered light distribution.

\subsection{Testing approximations}
\label{sec: approximations}
In Sect. \ref{sec: model}, we introduce several approximations that need to be tested. In this section, we evaluate the impact of dust settling. In Sect. \ref{sec: mcfost}, we test the single scattering approximation. A test of the isotropic scattering is given in Appendix \ref{sec: appendix-non-isotropic} and it is found to be irrelevant for the analysis conducted here. Finally, we show that the isothermal approximation is not appropriate in Appendix \ref{sec:vert_temp_gradient}.
\label{sec:settling}

Here, we focus on testing whether accounting for dust settling would affect the scattering height. The vertical distribution of the dust in protoplanetary disks is set by the balance between gravity, which pulls the grains towards the mid-plane, and turbulent diffusion, that stirs them up. Small grains that are more coupled to the gas are diffused more efficiently, while bigger grains, which are less coupled to the gas, are settled towards the mid-plane \citep{Birnstiel_2024, Villenave_2020}.
We treated settling following \citet{Dullemond_Dominik_2004}, who state that the gas to dust ratio $g(m,z)$ for a particles of mass, $m$, as a function of height can be computed integrating the following equation, namely,
\begin{equation}
    \rho_{\text{gas}}\frac{\partial g(m,z)}{\partial t}-D(m,z)\frac {\partial g(m,z)}{\partial z}+g(m,z)v_{\text{settle}}(m,z)=0,
\end{equation}
where, in the stationary case, this can be reduced to 
\begin{equation}  g(m,z)=g_0(m,z)\exp\left[\int_0^z\frac{v_{\text{settle}}(m,z')}{D(m,z')}dz'\right],
\label{eq: settling}
\end{equation}
with the settling velocity \citep{Birnstiel_2024} via
\begin{equation}
    v_{\text{settle}}(m,z) = -\frac{3}{4}\sqrt{\Omega_k}\frac{z}{\rho <v_{th}>}\frac{m}{\sigma}.
\end{equation}
$D(m,z)$ is the diffusion coefficient \citep{Youdin_Lithwick_2007} via
\begin{equation}
    D(m,z) = \frac{\alpha \ <v_{th}> H}{1+\text{St}^2},
\end{equation}
where $<v_{th}>$ is the mean thermal velocity, $\alpha$ is the  \citep{Shakura_Sunyaev_1973} parameter describing the magnitude of turbulence, and $\text{St}$ is the Stokes number, which in the Epstein regime \citep{Epstein_1924} is
\begin{equation}
    \text{St}=\frac{3}{4}\frac{m}{\sigma}\frac{\Omega_k}{<v_{th}>}\frac{1}{\rho}.
\end{equation}
All the quantities involved depend on the radius, so that the dust to gas ratio, $g(m,z)$, is a function of radius.
\begin{figure}
    \centering
    \includegraphics[width=\hsize]{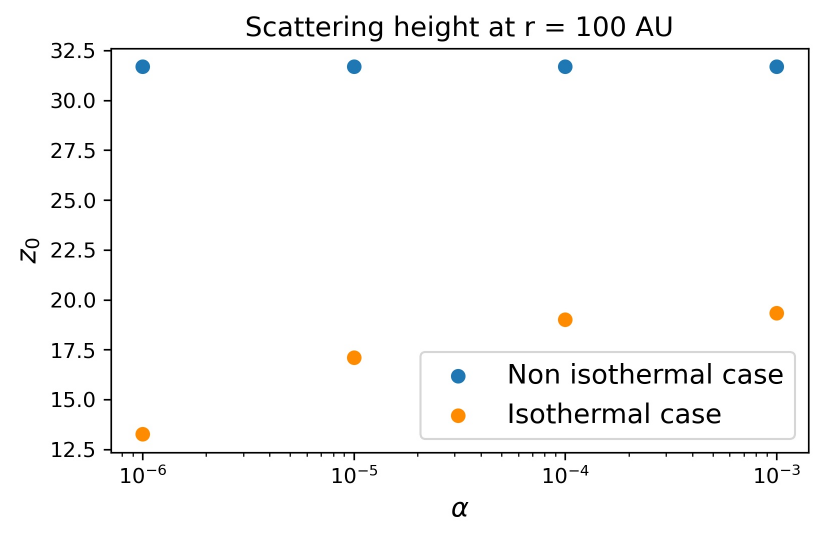}
    \caption{Scattering height at $r=100$ AU for different $\alpha$ parameters in the isothermal and non isothermal case, taking dust settling into account.}
    \label{fig:settling}
\end{figure}
We computed the scattering surface taking settling into account (i.e. weighting the dust densities with $g(m,z)$) in the isothermal and nonisothermal case, for grains with $a = 1\ \mu$m, using $m = 4\pi/3 \ \rho_s a^3$ and $\rho_s = 1.67  \ \text{g}/\text{cm}^3$. The results are shown in Fig. \ref{fig:settling}, where the scattering height at $r = 100$ AU as a function of $\alpha$ (which changes the settling efficiency).
We chose to test also the vertically isothermal model because \citet{Dullemond_Dominik_2004} found that in the isothermal case settling lowers the scattering surface and we find the same for very low values of the viscosity such as $\alpha=10^{-6}$, even $\mu$m grains are settled. In the nonisothermal case; however, we find that the scattering surface is independent of $\alpha$ and therefore settling does not influence the scattering surface. This is because the height at which settling becomes important occurs higher up than the scattering surface. To understand this, we look to Eq. \ref{eq: settling}.
The change in the dust to gas ratio $g(z)$ is due to the ratio between the settling velocity and the diffusion coefficient; since $v_{\text{settle}}\propto \rho^{-1}$ and $D\propto\rho$, the whole factor is proportional to $\rho^{-2}$. The density profile in the nonisothermal case is more vertically extended (a consequence of the higher temperature in the upper layers), as shown in the top panel of Fig. \ref{fig:isotermo_vs_non}, so that settling is less important. 
We tested the effect of settling for the other disks with an observationally measured temperature structure for the sample of \citet{exoALMA_temperature} and we conclude that $1 \ \mu$m grains are not settled in scattered light observations, as is typically assumed in observational studies \citep{Villenave2019}. Therefore, we find that ignoring settling when computing the scattering surface is a reasonable approximation.

\section{Comparison with the radiative transfer code \texttt{MCFOST}}
\label{sec: mcfost}
To assess the validity of the simplifying assumptions underlying our semi-analytic model, in particular, single scattering, we compared our results against the radiative transfer code \texttt{MCFOST} \citep{Pinte_2006, Pinte_2009}. This comparison serves as an independent validation of our approach, allowing us to identify the regime of inclinations where our model remains reliable. For simplicity, we compared the numerical radiative transfer results with an isothermal model, as the important physics is the same. We set up a disk model with a tapered profile, as in Eq. \ref{eq: surface_density_profile}, and a scale height of
\begin{equation}
    H(r)=0.1 \left( \frac{r}{100} \right)^{0.25}.
\end{equation}
We used astronomical silicates, with the opacities given Fig. 3 in \citet{Draine_Lee_1984}, where that of the J band ($\lambda \approx 1.2 \ \mu\hbox{m}$) does not diverge from the one adopted in our semi-analytic model (see Appendix \ref{appendix-opacity}). The grain size distribution as the same as in \citet{Mathis_1997} $n(a)\propto a^{-3.5}$ and with a maximum grain size of 1 mm. We ignore dust settling and consider isotropic scattering. We ray-traced the image at $\lambda = 1.2 \ \mu$m.

First we ran the model and used the command \texttt{-optical\_depth\_to\_cell} to extract the  optical depth from the star to every point in the disk. We find that the $\tau_1\approx 1$ surface in our model is consistent with the one computed with \texttt{MCFOST}. To obtain the scattering height from the computed images, we developed a simple algorithm similar to what is seen in observations (see, e.g., \citealt{Ginski_2016, DeBoer2021, Avenhaus2018}), but also applicable to smooth disks that do not show rings. In this case, we expect the isophotes to have the same role as ring structures; in fact, since we are considering isotropic scattering, the brightness of a point depends only on its distance from the star, and therefore each isophote traces a single radius. We fit ellipses to the isophotes and in this way we can measure the offset $u$ from the ellipses’ centre to the star and
then their height,
\begin{equation}
    z = \frac{u}{\sin i},
\end{equation}
with $i$ being the disk's inclination.

\begin{figure}[h!]
    \centering
  \includegraphics[width=\hsize]{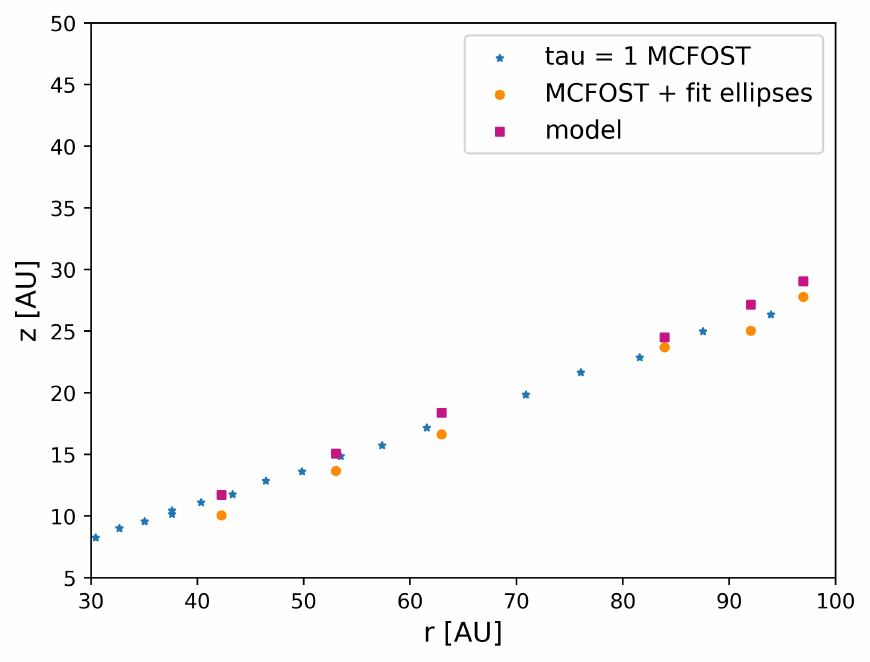}
    \caption{Comparison between the scattering surface from our semi-analytic model, the one computed with \texttt{MCFOST}, and the $\tau_1=1$ surface from \texttt{MCFOST}, at an inclination of $i=40^\circ$.}
    \label{fig:mcfost}
\end{figure}

We find that the scattering surface computed with \texttt{MCFOST} and ellipses fitting is comparable with the one of our model and with \texttt{MCFOST}'s $\tau_1\approx1$ surface, as shown in Fig. \ref{fig:mcfost}, serving as an empirical validation of our method.

\begin{figure}[h!]
    \centering
  \includegraphics[width=\hsize]{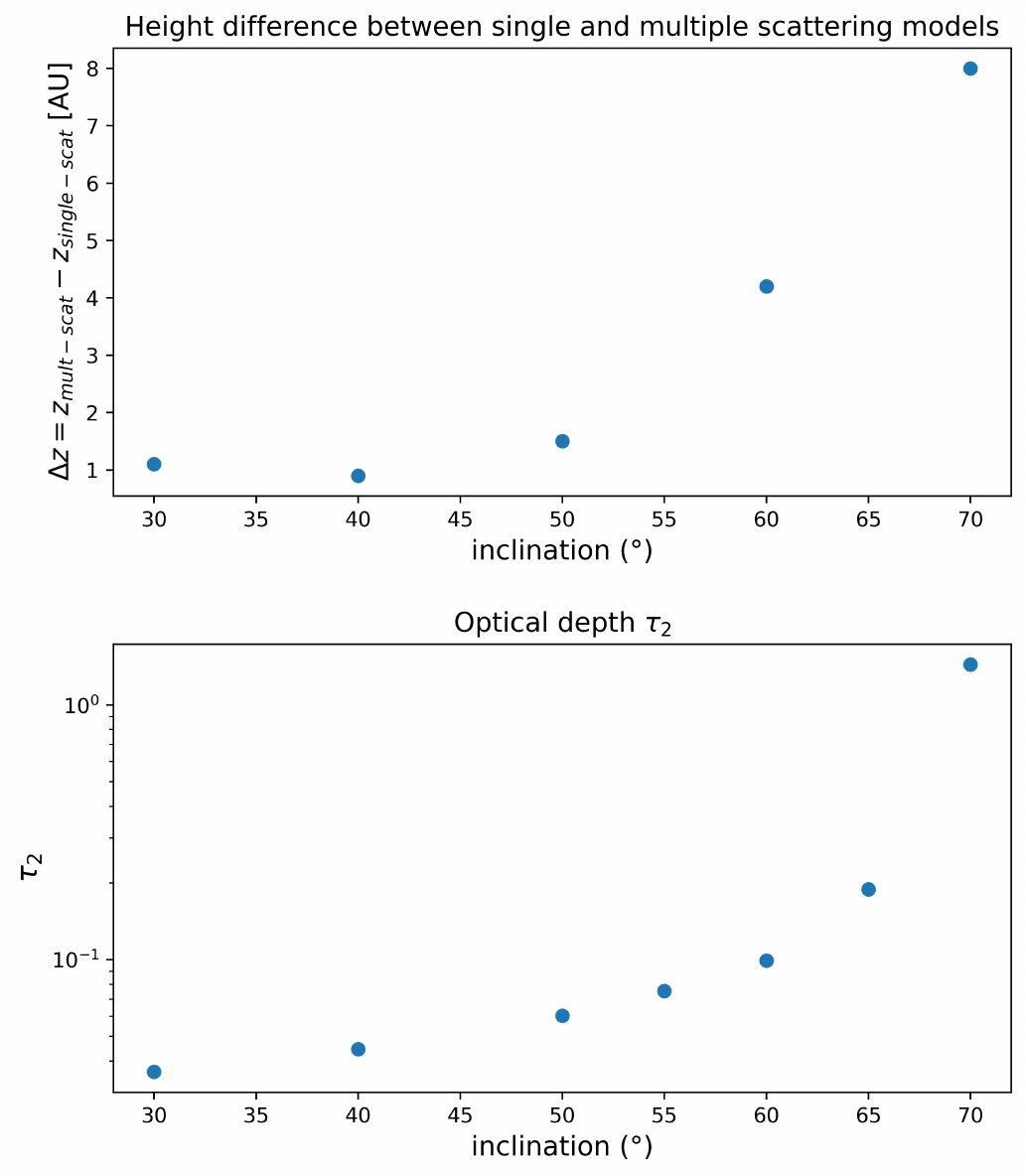}
    \caption{\textit{Top}: Difference in height between the scattering surface computed in the single scattering approximation or accounting for multiple scattering, at 200 AU. \textit{Bottom}: Optical depth from the scattering point to the observer as a function of disk's inclination.}
    \label{fig:tau_2}
\end{figure}

Furthermore, we tested the single scattering hypothesis, which is a critical one and one that makes the problem treatable semi-analytically. To include multiple scattering, we used the radiative transfer numerical model \texttt{MCFOST} turning on and off multiple scattering. We computed the scattering surface fitting the isophotes with ellipses as before. We find that for low inclinations, $i\leq 60^{\circ}$, multiple scattering does not affect the scattering height. For higher inclinations, however, the scattering surface is shifted to higher $z$. This behavior can be understood in terms of the optical depth structure of the disk. The location of the scattering surface is defined by $\tau_1 = 1$, which marks where stellar photons undergo their first interaction with the disk surface. Since regions with $\tau_1 > 1$ are optically thick to the incoming stellar radiation, scattering events occurring deeper in the disk do not alter the position of the surface. When multiple scattering is accounted for, however, the optical depth to a second scattering, $\tau_2$, becomes relevant. Once the condition $\tau_2 \ll 1$ no longer holds, additional scattering events occur, producing a net upward shift of the effective scattering surface. Consequently, the inclination at which multiple scattering becomes nonnegligible coincides with the breakdown of the $\tau_2 \ll 1$ assumption, which in our models occurs at approximately $i \simeq 60^\circ$, Fig. \ref{fig:tau_2} shows an example: we can see that as the inclination increases $\tau_2$ becomes higher (bottom panel), and consequently the difference between the scattering height computed with the single or multiple scattering approximations increases as well (top panel).
We point out that our analysis does not account for the case in which observations are performed in polarized light. A more detailed study should consider radiative transfer for scattered polarized light. However, we expect that the dominant contribution to the polarized signal comes from single scattering, as multiple scattering strongly reduces the polarization fraction. This is consistent with our approach, since our model is valid precisely in the single scattering regime, and in fact, observations in polarized light naturally select this regime, as the multiple scattering contribution is suppressed.

\section{Results}
\label{sec: results}
Section \ref{sec: methods} presents the direct problem: we determined the scattering height based on the relevant disk parameters. When dealing with observations, however, we are typically more interested in the inverse problem, namely to infer the disk parameters from the observed scattering surface. This is the problem we consider in this section. In particular, from the observed scattering height, we aim retrieve the mass in small ($\mu$m sized) dust. Inverting Eq. \ref{eq: semi-analytical model}, we obtain
\begin{multline}
    \kappa M_{sd}(R) = (2\pi)^{3/2} r_c \, (1-e^{-R/r_c}) \, \sqrt{k_b} \, (\mu m_H G M_*)^{-1/2} \cdot \\
    \cdot
         \left(\int_0^{s_0} 
            \frac{\sqrt{T_{mid}(r)}}{T(z,r) \, r^{5/2}} 
            e^{-r/r_c}  
            e^{-\frac{\mu m_H G M_*}{k_b} \int_0^z \frac{z' dz'}{(r^2+z'^2)^{3/2} T(z',r)}} 
          \, ds\right)^{-1}.
    \label{eq: inverse}
\end{multline}
Here, the cumulative mass is contained in small dust up to the radius, R; we note that there is a degeneracy between the small dust mass, $M_{sd}$, and the opacity, $\kappa$. Some considerations on opacity models and uncertainties are in Appendix \ref{appendix-opacity}. Additionally, we point out that this estimate assume a radially smooth distribution of small dust grains, without accounting for gaps or rings.
\subsection{Sample}
\label{sec: sample}
We now wish to build an observational sample of disks we can apply our model to. To apply Eq. \ref{eq: inverse}, we need the following observables:
\begin{itemize}
    \item  coordinates of the scattering surface $(r_0,z_0)$, inferred from the morphology of the rings in scattered light;
    \item  thermal structure of the disk, which is derived from spatially resolved observations of CO rotational lines;
    \item  stellar mass;
    \item characteristic radius, $r_c$, that marks the onset of the exponential taper in the surface density profile. In our sources this parameter is constrained by fitting rotation curves.

\end{itemize}
Therefore, constraints on the properties of small dust grains in protoplanetary disks require multi-wavelength observational data that include scattered light imaging and molecular line emission. From the literature, we were able to find ten disks that have all the required observations. The main limitation consists in having the first two.  Table \ref{table:1} reports the references for the literature measurements of scattering heights, temperature structures and stellar masses, for our disk sample.  The cut-off radius $r_c$ has been taken from \citet{Martire_2024} and \citet{exoALMA_rotation}, however, this parameter does not have a big impact on our results. The top panel of Fig. \ref{fig:osservazioni scremate} reports the measured scattering surface from the literature for our sample. The typical uncertainties in the scattering height measurements are on the order of 10-30\%, with some outliers \citep{Byrne_2026}; we discuss the temperature data in Sect. \ref{sec: thermal structure}.

\begin{table*}[h!]
\caption{Sample with references for scattering heights, temperatures, and stellar masses.}                
\label{table:1}    
\centering                        
\begin{tabular}{c c c c}      
\hline\hline               
Disk & Temperature & Scattering height & Stellar mass\\        
\hline                      
IM Lup & \citet{Law_2021} & \citet{Avenhaus2018} &  \citet{Avenhaus2018}\\
V4046 Sgr & \citet{exoALMA_temperature} & \citet{Avenhaus2018} &  \citet{Avenhaus2018}\\
HD 163296 & \citet{Law_2021} & \citet{Ginski2023} &  \citet{Mass_HD163296}\\
LkCa15 & \citet{exoALMA_temperature} & \citet{Ginski2023} & \citet{Law_2023} \\
RX J1615 & \citet{exoALMA_temperature} & \citet{Avenhaus2018} & \citet{Avenhaus2018}\\
MWC 480 & \citet{Law_2021} & \citet{Roumesy_2025} & \citet{Flaherty2020} \\
SY Cha & \citet{exoALMA_temperature} & \citet{Ginski_2024} & \citet{Ginski_2024} \\
J1852 & \citet{exoALMA_temperature} & \citet{Ginski2023} & \citet{Villenave2019} \\
HD 97048 & \citet{Pezzotta_2026} & \citet{Ginski_2016} & \citet{Ginski_2016} \\
GM Aur & \citet{Law_2021} & \citet{Byrne_2026} & \citet{Maps_Teague} \\

\hline\hline
\end{tabular}
\end{table*}

\subsection{Mass in small grains}
We derived the small dust mass for the sample described in Sect. \ref{sec: sample}. We computed the small dust masses applying Eq. \ref{eq: inverse} with $\kappa = 10^4 \text{ cm}^2/\text{g}$ (see Sect. \ref{sec: model} and Appendix \ref{appendix-opacity}). Since the expression for the mass involves integrals that must be evaluated numerically, a standard error propagation is not straightforward. To estimate the uncertainty on the derived mass, we randomly perturbed the input parameters within their respective observational uncertainties and recompute the resulting $M_{sd}$ values for each realization. The uncertainty on the mass is then taken as the standard deviation of the resulting distribution in logarithmic space, namely, $\log(M_{sd})$. It is important to point out that the scattering surface data points correspond to ring locations; therefore, with our model we estimate the cumulative mass for grains with $a\leq 1\ \mu$m up to the ring where the scattering height has been measured. When multiple rings are present, we treat each ring independently, that is, we computed the small dust mass enclosed within each ring without subtracting the contribution from inner rings. This is particularly evident in the case of HD~97048, where the scattering surface is not a monotonic function of radius and consequently neither is the inferred cumulative mass.

\begin{figure}
    \centering
    \includegraphics[width=\hsize]{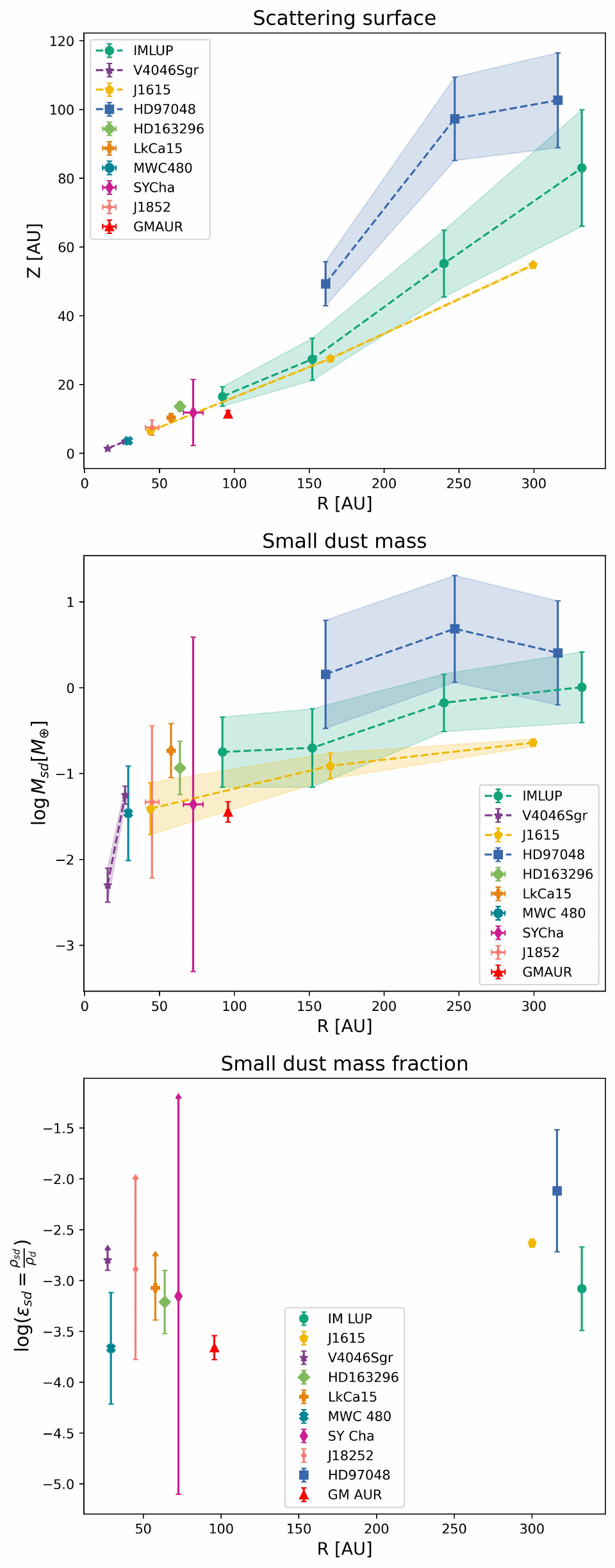}
    \caption{\textit{Top}: Measured scattering surfaces from the literature. \textit{Middle}: Small dust masses as a function of radius inferred from the inverse problem in our work. \textit{Bottom}: Small dust mass fraction for all the disks in the sample.}
    \label{fig:osservazioni scremate}
\end{figure}

The results are presented in Fig. \ref{fig:osservazioni scremate}: in the top panel, there are the measured scattering surfaces from the literature and in the middle panel, our measurements are given as a function of radius. The small dust masses are also reported in Table \ref{table:2}. 

From these results, the following considerations can be made: most of the masses are in the range $10^{-2}M_{\oplus} \leq M \leq 1 M_{\oplus}$ and the disks with more than one ring in general exhibit an increasing trend of the mass with the radius (with the previously mentioned exception of HD 97048); this is expected because the measured mass is the cumulative mass up to the ring radius. SY Cha has a particularly large uncertainty in its small dust mass estimate. This is primarily due to the uncertainty on the location of the scattering surface, which dominates the overall error. Finally, even if HD 97048 has a high and flared scattering surface, its mass is comparable to the other disks’ mass, and the geometry of the surface can be explained by his unusually high temperature \citep{Pezzotta_2026}.

\begin{table}
\caption{Small dust mass measurements, where measurement refers to the cumulative mass at the radius of the given ring.}    
\label{table:2}    
\centering                        
\begin{tabular}{c c c}      
\hline\hline               
Disk  & Radius (AU) & Mass ($\log(M_{sd} \ [M_{\oplus}])$)  \\
\hline
V4046 Sgr ring 1 & $15.35\pm0.06$ &  $-2.30 \pm 0.2$ \\
V4046 Sgr ring 2 & $27.01\pm 0.10$ & $-1.25\pm 0.10$ \\
RXJ 1615 ring 1  & $44.00\pm 0.26$ & $-1.41\pm 0.30$ \\
RXJ 1615 ring 2  & $164.00\pm 0.54$ & $-0.91\pm 0.15$ \\
RXJ 1615 ring 3  & $229.44\pm 1.99$ & $-0.64\pm 0.04$ \\
IM Lup ring 1    & $91.90\pm 3.17$  & $-0.75\pm 0.41$ \\
IM Lup ring 2    & $152.11\pm 4.75$ & $-0.70\pm 0.46$ \\
IM Lup ring 3    & $240.84\pm 4.75$ & $-0.17\pm 0.33$ \\
IM Lup ring 4    & $332.75\pm 12.68$ & $0.01\pm 0.41$ \\
HD 163296 & $63.6\pm 1.58$ &  $-0.93\pm 0.31$ \\
LkCa 15   & $57.7\pm1.63$ &  $-0.73 \pm 0.32$ \\
MWC 480   & $29.1\pm 0.99$ & $-1.46\pm 0.55$ \\
J1852   & $147.1\pm3.41$ &  $-1.33\pm 0.89$ \\
SY Cha & $72.5\pm 6.6$ & $-1.36\pm 1.95$ \\
HD 97048 ring 1 & $160.8\pm16.9$ & $0.16\pm 0.63$  \\
HD 97048 ring 2 & $247.1\pm28$ & $0.69\pm 0.62$  \\
HD 97048 ring 3 & $340.6\pm34.8$ & $0.41\pm 0.61$  \\
GM Aur & $104.78 \pm 1.25$& $-1.45\pm 0.19$\\
\hline\hline
\end{tabular}
\end{table}

\subsection{Small dust ratio}
\label{sec: small dust ratio}
We now try to give additional constraints on the dust distribution combining our results with data from continuum submillimeter (submm) observations. Large grains dominate the mass budget of the grain size distribution; therefore, the dust mass estimated from the continuum sub-mm data can be considered as the total dust mass, $M_d$. We computed the small dust ratio as 
\begin{equation}
    \epsilon_{sd}=\frac{M_{sd}}{M_d}.
\end{equation}
It is important to highlight a key distinction: when measuring the small dust mass, we consider only the material within the radius of a given ring, while the large dust mass measurements are integrated over the entire disk. The ideal scenario would be to have very outer rings to capture the total small dust mass, $\epsilon_{sd}$, or alternatively, to have radius-dependent large dust mass measurements. The latter is only available for the MAPS sample (IM Lup, GM Aur, HD 163296, MWC 480), that has multi-wavelength data, and for which \citet{Sierra_2021} were able to fit simultaneously for the surface density $\Sigma(r)$ and the maximum grain size $a_{\text{max}}(r)$. For these disks, we integrated the surface density profile obtaining the mass as a function of radius, 
\begin{equation}
    M(r)=\int_0^r\Sigma(r)2\pi r dr.
\end{equation} 
For the rest of the sample, we computed $\epsilon_{sd}$ based on the available total dust masses from \citet{Curone_2025, Walsh_2016}. These masses are computed under the optically thin assumption, using the equation,
\begin{equation}
    M_d = \frac{d^2 F_{\nu}}{\kappa_{\nu}B_{\nu}}
    \label{eq: optically_thin},
\end{equation}
with $\kappa_{\nu}=3.5 \text{ cm}^2\text{g}^{-1}  \ \cdot 870 \ \mu\text{m}/\lambda$ \citep{Beckwith_1990}. In cases where the scattered light measurements are limited to small-radius rings, the results may be underestimated and those cases are reported in Fig. \ref{fig:osservazioni scremate} with pointing up arrows on the error bar.
The small dust ratio (i.e. the mass fraction of micron-sized grains) computed for all the disks in the sample is reported in Table \ref{table:3} and shown in the bottom panel of Fig. \ref{fig:osservazioni scremate}, where it can be seen that the majority of the ratios are in the range $-3.5\leq \log(\epsilon_{sd})\leq -2.5$, and all the disks fall in the range $-4\leq \log(\epsilon_{sd})\leq -2$. We present a discussion of the information that can be extracted from these results in the following subsections.

\section{Discussion}
\label{sec: discussion}
\subsection{Comparison with exoALMA XV}
\citet{Rosotti_2025} recently developed a model for the CO emitting height, within the \textit{exoALMA} sample \citep{Teague_2025}, which is very similar to what we present here. They adopted a semi-analytical approach to link the CO emitting height to the disk temperature structure and gas column density, assuming a fixed CO abundance. In the model of \citet{Rosotti_2025}, the important quantity is the optical depth between the observer and the CO emitting height. This requires integrating vertically along the $z$ direction and makes the problem local, allowing them to infer the local gas surface density. In contrast, reflecting the different physical mechanisms underlying CO emission and scattered light, in our model the relevant optical depth, $\tau_1$, is integrated from the star to the scattering point (i.e. radially). As a consequence, it depends on the full radial and vertical structure of the disk. The problem is therefore nonlocal and our method does not allow us to derive a local surface density, but an integrated quantity corresponding to an enclosed mass instead.

\subsection{Impact on chemistry}
 
The abundance of small grains plays an important role in the chemical modeling of protoplanetary disks because it controls both the available surface area for grain-surface reactions and the penetration of UV radiation into the deeper layers of the disk. Since small grains dominate the total surface area, they enhance the adsorption of gas-phase molecules, the formation of ices and surface chemistry reactions, such as hydrogenation \citep{Aikawa_Nomura_2006}. At the same time, they dominate the opacity at at shorter wavelengths, so that a higher fraction of small grains increases UV shielding, modifying the photo-dissociation rates and the gas thermal balance \citep{Woitke_2016}. In general, a smaller amount of small grains in the disk atmosphere can alter the chemical abundances through the disk and lead to a chemically more active disk interior \citep{Vyasunin_2011}. The values we derived for the small dust mass ratio are valuable constraints for chemical models \citep[e.g., DALI;][]{ Bruderer_2012, Bruderer_2013} that typically take the small dust mass ratio as input.
\begin{table}
\caption{Small dust mass fraction, $\epsilon_{sd}$.}
\label{table:3}    
\centering   
\begin{threeparttable}
\begin{tabular}{c c c}      
\hline\hline               
Disk & Radius [AU] & Mass fraction $\log\epsilon_{sd}$\\        
\hline   
V4046 Sgr\tnote{a} & $27.01\pm 0.10$ & $-2.80\pm 0.10$ \\
RXJ 1615  & $229.44\pm 1.99$ & $-2.63\pm 0.04$ \\
IM Lup    & $332.75\pm 12.68$ & $-3.08\pm 0.41$ \\
HD 163296 & $63.6\pm 1.575$ &  $-3.21\pm 0.31$ \\
LkCa 15\tnote{a}   & $57.7\pm1.634$ &  $-3.07 \pm 0.32$ \\
MWC 480   & $29.1\pm 0.989$ & $-3.67\pm 0.55$ \\
J1852\tnote{a}  & $147.1\pm3.413$ &  $-2.89\pm 0.89$ \\
SY CHA \tnote{a}& $72.5\pm 6.6$ & $-3.15\pm 1.95$ \\
HD 97048 & $340.6\pm34.8$ & $-2.12\pm 0.61$  \\
GM Aur & $104.78 \pm 1.25$& $-3.65\pm 0.19$\\

\hline\hline
\end{tabular}
\begin{tablenotes}
\item[a] The small dust mass fraction is a lower limit
\end{tablenotes}
\end{threeparttable}
\end{table}

\subsection{Grain size distribution}
\label{sec: gsd}
We go on to interpret the small dust ratios we have obtained in light of typical grain size distributions used in the protoplanetary disk literature. In the interstellar medium, in debris disks or asteroids, dust follows an MRN distribution via $n(a)\propto a^{-q}$, with $q = 3.5$ \citep{Mathis_1997,Tanaka_1996}, whereas dust in
protoplanetary disks evolves under both coagulation and fragmentation, leading to more complex distributions. Some typical limit cases include fragmentation-limited distributions, with an exponent of $3.5$ \citep{Williams_1994}, drift-limited ones, with $q = 2.5$ \citep{Birnstiel_2012}, or even steeper distributions, produced when fragmentation is induced by radial drift rather than turbulence, with $q=4$ \citep{Birnstiel_2015}. Understanding these distributions is essential for interpreting observations, modeling dust evolution, and constraining planet formation processes.

To this point, we computed the small dust mass ratio using reference values of the opacity (i.e, a value of $10^4 \text{ cm}^2/\text{g}$ for the small dust and a value of $3.5 \text{ cm}^2/\text{g}$ for the large dust). However, in reality, we would expect the two opacities to depend on the properties of the grain size distribution (as well as on the dust composition and porosity). In practice, this means that the value we have found of the small dust ratio is not necessarily consistent with the opacity value we have assumed. To solve this problem, we attempted to use both the submm and scattered light constraint at the same time in a self-consistent manner. As we discuss in the following shortly, solutions consistent with the observational constraints are possible (for a given choice of dust composition and porosity) only for specific grain size distributions. Our results can therefore be interpreted as constraints on the grain size distribution. We detail our procedure to find a self-consistent solution below.

Given an assumed power-law grain size distribution $n(a)=c \ a^{-q}$, we defined the mass in small dust as the mass in grains with $a_{\text{min}}\leq a\leq a_{\text{thresh}}$ and the one in big grains as the mass in grains with $a_{\text{thresh}}\leq a \leq a_{\text{max}}$. The small dust ratio therefore becomes
\begin{equation}
    \epsilon_{sd}=\frac{a_{\text{thresh}}^{4-q} - a_{\text{min}}^{4-q}}{a_{\text{max}}^{4-q}-a_{\text{thresh}}^{4-q}}.
\end{equation}
We chose $a_{\text{thresh}}=1\ \mu$m to be the grain size that accounts for the 90\% of the opacity in the J band, as explained in Sect. \ref{sec: model}, and $a_{\text{min}}=0.1 \ \mu$m (noting that a change in this last parameter only has a minor effect on the results). We compute $\epsilon_{sd}$ for $10^{-4} \text{ cm} \leq a_{\text{max}}\leq 10^{3} \text{ cm}$ and $2\leq q \leq 4$, defining this as the true small dust ratio, $\epsilon_{\text{sd}}^{\text{true}}(a_{\text{max}},q)$. 

As previously explained, to compare $\epsilon_{\text{sd}}^{\text{true}}$ with the value derived from our measurements, we must account for the opacity assumptions underlying both our model and literature submm measurements. Since the opacity depends on the grain size distribution, we have to recompute the opacity used in our measurements for each grain size distribution. Thus, rather than using reference values for the opacities, we computed a measured small dust ratio, $\epsilon_{\text{sd}}^{\text{measured}}(a_{\text{max}}, q)$, for each grain size distribution, consistent with the opacities used in each case.

For each disk in our sample, given a fixed dust composition, we recomputed the dust mass as follows. Using \texttt{Optool}, we calculated the opacity for $a_{\text{max}} = 1~\mu$m, while varying $q$ between 2 and 4. We then derived $M_{\text{sd}}$ via Eq.~\ref{eq: inverse}. For the large dust mass, we started from the literature values \citep{Martire_2024, Curone_2025, Walsh_2016}, computed under the optically thin assumption, with Eq. \ref{eq: optically_thin}.
We then recomputed the opacity, $\kappa_{\nu}'$, with \texttt{Optool} for each pair $(a_{\text{max}}, q)$ and obtained a grid of large dust masses as
\begin{equation}
M_{d} = M  \frac{\kappa_{\nu}}{\kappa_{\nu}'}.
\end{equation}
This allowed us to derive $\epsilon_{\text{sd}}^{\text{measured}}(a_{\text{max}}, q)$ self-consistently with the choice of the opacity.

Finally, we compared $\epsilon_{\text{sd}}^{\text{true}}$ and $\epsilon_{\text{sd}}^{\text{measured}}$. Only models where the two values are the same are acceptable and, thus, this allows us to evaluate
\begin{equation}
f=\frac{\epsilon_{\text{sd}}^{\text{measured}}}{\epsilon_{\text{sd}}^{\text{true}}}
\end{equation}
and determine the regions in the $(a_{\text{max}}, q)$ plane where $f = 1$, thereby constraining the grain size distribution.
\begin{figure}
    \centering
    \includegraphics[width=\hsize]{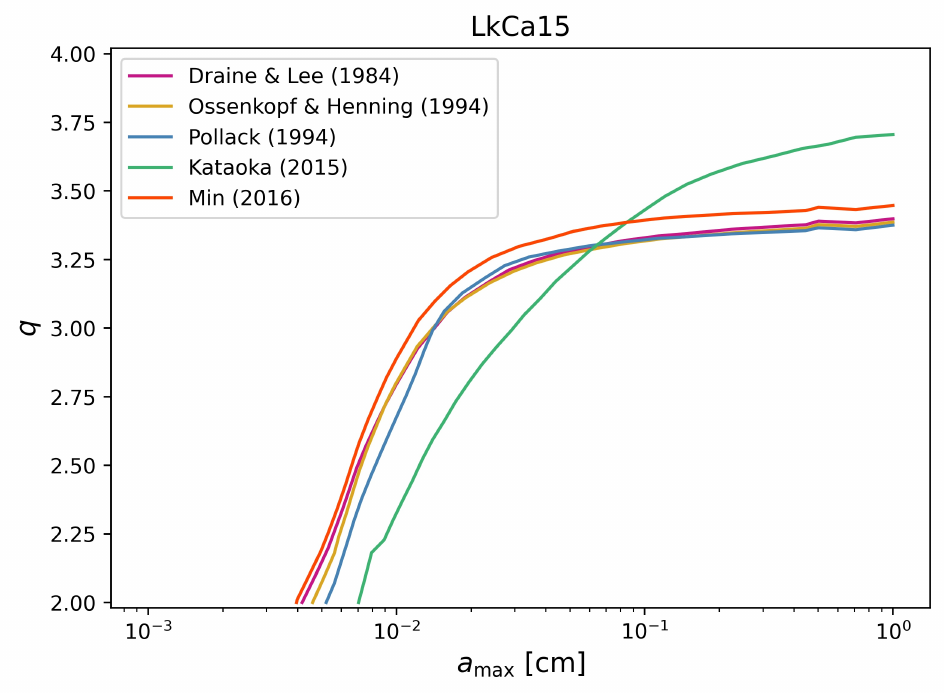}
    \caption{Constraints on the maximum grain size and power-law exponent for LkCa15 obtained using different opacity models.}
    \label{fig: LkCa15_contours}
\end{figure}

Figure~\ref{fig: LkCa15_contours} shows the resulting constraints for LkCa15 under various opacity models \citep{Draine_Lee_1984, Pollack_1994, Ossenkopf_Henning_1994, Min_2016, Kataoka_2015}. These models differ in terms of both the composition and porosity: the models by \citet{Draine_Lee_1984} and \citet{Pollack_1994} represent compact grains made of astronomical silicates and carbon, while the model by \citet{Ossenkopf_Henning_1994} accounts also for thin icy mantles. The model from \citet{Min_2016} represents grains with moderate (30\%) porosity, while the model by \citet{Kataoka_2015} represents highly porous grains with $p=90\%$. The corresponding plots for other disks are presented in Appendix~\ref{appendix-gsd}. 

Small maximum grain sizes ($a_{\text{max}}<100\mu$m) are not compatible with typical exponents. Our results then imply that some level of grain growth is necessary to simultaneoulsy interpret submm and scattered light observations. This is consistent with multiwavelenght results at radio and mm wavelengths, but we remark that in this case the constraint is provided by a different wavelength and methodology. For some disks, values of $q \approx 3.5$ are difficult to reproduce across all opacity models, implying a shallower grain size distribution, with the exception of the \citet{Kataoka_2015} model, which assumes highly porous grains. While composition has little effect on the $f=1$ contour, porosity plays an important role.
\begin{figure}
    \centering
    \includegraphics[width=\hsize]{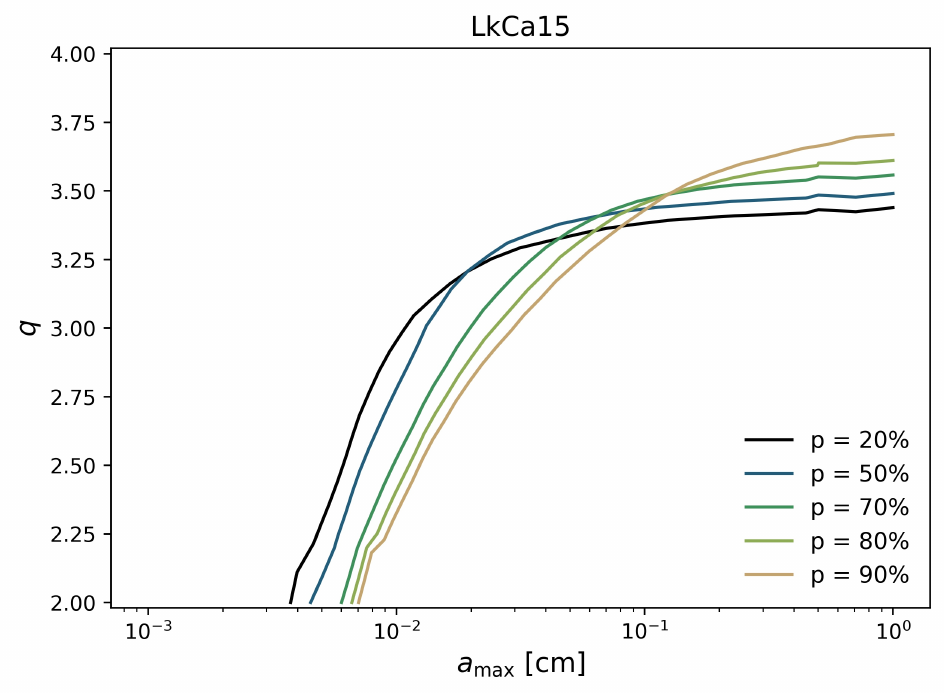}
    \caption{Constraints on the maximum grain size and power-law exponent for LkCa15 using the \citet{Pollack_1994} composition and varying porosity.}
    \label{fig: LkCa15_composition}
\end{figure}
\begin{figure}
    \centering
    \includegraphics[width=\hsize]{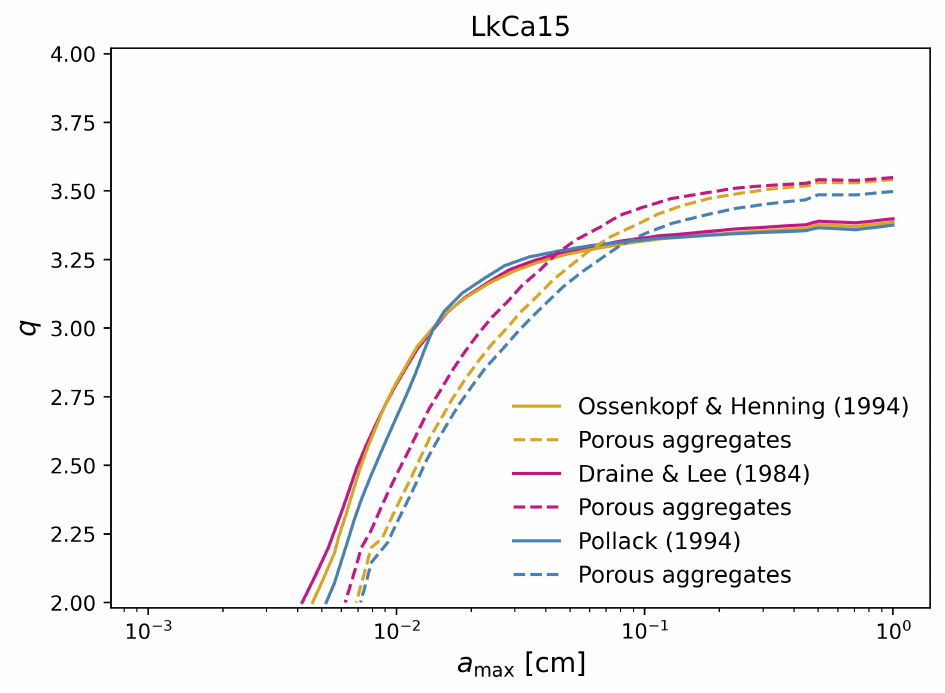}
    \caption{Constraints on the maximum grain size and power-law exponent for LkCa15 using \citet{Draine_Lee_1984}, \citet{Pollack_1994}, and \citet{Ossenkopf_Henning_1994} opacities. The dotted lines represent the same composition, but considering a porous aggregates scenario, as big grains have a high porosity ($p\sim0.8$).}
    \label{fig:porous_aggregates}
\end{figure}

Figure~\ref{fig: LkCa15_composition} illustrates the impact of porosity on the $f=1$ contour for the \citet{Pollack_1994} composition. The variation in porosity, significantly alters the contour shape, introducing a major source of uncertainty. 
A similar effect is seen in Fig. \ref{fig:porous_aggregates}, where we compare a grain population composed of compact small grains and highly porous large aggregates with the corresponding zero-porosity case. In both setups, porosity significantly impacts the inferred constraints.
However, porosity can be constrained observationally through the dust scattering phase function, which encodes the dust grain properties, especially the porosity and the fractal dimension of the aggregates \citep{Tazaki_2019}. The phase function, derived from the intensity of scattered light as a function of angle, can be inferred from the measured scattering height and brightness distribution. Such analyses have been performed for individual disks \citep{Ginski_2016, Tazaki_2023, Columba_2025} and for larger samples \citep{Ginski2023}. A similar analysis is beyond the scope of this work, but we point out that the uncertainty due to the porosity can be reduced in this way.

Finally, we highlight a few caveats of this analysis: we considered a single opacity for either the big or the small grain populations, without considering variations in the vertical or radial opacity. 
Additionally, these constraints refer to the disk-integrated small-dust ratio. They do not capture radial variations, and substructures within the disk may host grain-size distributions that differ substantially from the global average. Finally, we assume that the small and large grain populations follow the same underlying size distribution, which allows us to directly compare their properties and draw conclusions on the global grain size distribution. While in reality the two populations may evolve independently, relaxing this assumption would make the comparison between the two populations ill-defined, preventing any constraint on the overall grain size distribution.

\subsection{The importance of the vertical temperature gradient}
\label{sec: thermal structure}
It is common when modeling proto-planetary disks to assume that they are vertically isothermal. Our results, detailed in Appendix \ref{sec:vert_temp_gradient} and Sect. \ref{sec:settling} show that, at least as far as the scattered light height is concerned, such an assumption has a significant impact on the results. On the one hand, including the vertical temperature gradient pushes the scattering surface significantly higher up than in an isothermal model; on the other hand, it also makes the effect of dust settling unimportant. As a consequence, in this paper we were only able to apply our method to a rather limited disk sample, employing the strong constraint that each individual disk needs to have a measured temperature profile from molecular submm emission, using, for example, the method from \citet{Law_2021}. There is indeed a considerable difference from source to source in the parameters describing how the temperature varies with the vertical coordinate and one cannot simply extrapolate from the midplane temperature. If more observations of this kind become available in the future, it will enable us to extend our analysis to a larger sample.

It should also be kept in mind that even though we have taken the temperature profiles measured by CO ALMA observations as input for the purposes of this work, there is considerable uncertainty associated to them. Fundamentally, this comes from the fact that the emission of a given transition is associated with a single emitting layer and, therefore, at each radial distance, it only provides a temperature measurement at a single height. The more transitions are available$-$following the approach of \citealt{Pezzotta}, who employed seven different transitions to extract multiple rotation curves in the disk around HD163296, and extending it to also measuring the temperature$-$the better one can allow us to measure how the temperature varies with the vertical coordinate. However, given that most existing measurements are based only on two transitions (most often $^{12}$CO and $^{13}$CO), there is a significant amount of extrapolation in the profiles reported in the literature. In addition, it is likely that the temperature structure of real disks is  more complex than the simple radial power-law behavior often assumed, as recently shown by \citet{FehrAndrews} through a nonparametric method and as can be inferred directly using tomography on edge-on disks (\citet{Dutrey_2017, Flores_2021} as well as the upcoming results from the DiskStrat ALMA large program).

\subsection{Comparison with literature models}

Radiative transfer models of protoplanetary disks in scattered light are already available in the literature. For instance, \citet{Villenave2019} used \texttt{MCFOST} to model SPHERE observations of J1852 and J1608, analyzing the different radial extents of small and large dust grains. Similarly, \citet{Muro-Arena_2018} reproduced the ALMA and SPHERE observations of HD~163296 with the radiative transfer code \texttt{MCMax3D} \citep{Min_2009} to investigate the distinct radial morphologies of the disk in the millimeter and NIR regimes. \citet{Rich_2021} focused instead on the relative heights of the scattering surface and the CO emission surface, using them to constrain the dust-to-gas ratio in HD 163296, HD 97048, and IM Lup. 
These studies differ from ours because the dust properties were fixed; in particular, the grain size distribution was assumed a priori, while other physical parameters such as the spatial distribution of grains were varied to reproduce the observations. However, from the assumed grain size distribution, the small dust ratio is 0.08 for HD 163296 and 0.02 for J1852, higher and not compatible with what we found. By contrast, \citet{Franceschi_2023} employed \texttt{RADMC-3D}\footnote{\url{https://www.ita.uni-heidelberg.de/dullemond/software/radmc-3d/}} to constrain the turbulence and the vertical and radial distribution of large and small grains in IM Lup, fitting for the grain size distribution. They found that reproducing the scattered-light emission required a mass fraction of micron-sized grains of $\sim10^{-4}$, slightly lower than our median value of $10^{-3.5\pm0.45}$ between the rings, but still comparable to our result. Recently, \citet{Swastik_2025} modeled LkCa15 simultaneously in the millimeter and scattered-light regimes, constraining the scattering phase function as part of the fit. They found that the model that best matches the observations is characterized by highly porous aggregates, with a porosity p= 87\% and a local grain-size distribution exponent of 2.3 between 12 $\mu$m and 2 mm. We can compare this with what we found for LkCa15, considering Fig. \ref{fig:porous_aggregates} we find that for $q = 2.3$ and $p=90\%$ this corresponds to $a_{\text{max}}\approx 5 \hbox{ mm}$, which is comparable with the result of \citet{Swastik_2025}.

It is worth highlighting that our approach differs from those presented in the cited works because we have focused solely on the height of the scattering surface, using a semi-analytical modeling framework, rather than on reproducing in detail the intensity of the emission. Accepting this downside, our method does not require detailed modeling of individual disks and, therefore, it can be efficiently extended to larger samples.

\section{Conclusions}
\label{sec: conclusions}
We used scattered-light surface height measurements to directly constrain the mass of small dust grains in protoplanetary disks. To do so, we developed a semi analytical model for the scattering surface. Starting from radiative transfer considerations, we modeled the path of the stellar radiation and its interaction with the disk to understand what is the basic physics that sets the height of the surface. The results of our theoretical investigation are as follows:
\begin{itemize}
    \item The scattering surface coincides with the $\tau_*=1$ surface, and the scattering height can be expressed as a function of the disk's parameters, $z_0(r_0,\kappa,M_{\text{small dust}},T(r,z))$. This provides a direct link between observable surface heights and physical disk parameters.
    \item The determination of the scattering surface is a nonlocal problem. The height of the surface at a given radius depends not only on the local column density, but also on the integrated density along the entire optical path from the star to the scattering point.
    \item We tested this prediction against the radiative transfer model \texttt{MCFOST} and consistency was found.
    \item When considering disks in scattered light, the isothermal approximation is not correct. Ignoring the temperature gradient (a common simplification) leads to significant underestimation of the surface height ($>30\%$). This necessitates spatially resolved temperature measurements for accurate mass estimates.
    \item In contrast, dust settling and anisotropic scattering have negligible effects on the scattering surface location and can be safely ignored. Multiple scattering becomes important only for high inclinations ($i\geq60^{\circ}$).
\end{itemize}

We then approached the inverse problem: using measured scattering surface heights to constrain the properties of small dust grains ($a\leq 1 \ \mu$m). The inversion is degenerate in terms of dust opacity and mass, so by assuming an opacity model, we were able to estimate the small dust mass for a sample of ten protoplanetary disks. From these measurements, we were able to obtain the following results. 
\begin{itemize}
    \item We used the combined scattered-light data (probing small grains) with submm data (probing large grains) to derive the small dust ratio: the fraction of small dust mass- finding typical values around $10^{-3}$;
    \item By requiring self-consistency between scattered-light (small dust) and submm (total dust) observations, we constrained the grain size distribution. For moderate grain growth $a_{\text{max}}\geq 0.1 \text{ mm}$, we retrieved common power-law indices of $q \sim 3-3.5$.
    \item Dust porosity significantly affects these constraints and represents a major uncertainty. This can be reduced through scattering phase function analyses, which independently constrain porosity and grain structure, as in \citet{Tazaki_2023, Ginski2023}. 
\end{itemize}
Taken together, these results show that scattered-light surface heights, combined with thermal structure measurements, can provide constraints on the small dust grain population in protoplanetary disks, in particular on its mass. By combining this diagnostic with submm data, one can obtain constraints on the grain size distribution. While porosity remains one of the major sources of uncertainty, this method, combined with scattering phase function analyses, can provide a statistical characterization of the small dust population in disks.

\begin{acknowledgements}
We thank Anibal Sierra for kindly providing us with the MAPS fits used in this work. We acknowledge support from the European Union (ERC Starting Grant DiscEvol, project number 101039651) and from Fondazione Cariplo, grant No. 2022-1217. MB has received funding from the European Research Council (ERC) under the European Union’s Horizon 2020 research and innovation programme (PROTOPLANETS, grant agreement No. 101002188). S.F. acknowledges financial contributions from the European Union (ERC, UNVEIL, 101076613) and from PRIN-MUR 2022YP5ACE. Views and opinions expressed are, however, those of the author(s) only and do not necessarily reflect those of the European Union or the European Research Council. Neither the European Union nor the granting authority can be held responsible for them.
\end{acknowledgements}

\bibliographystyle{aa}
\bibliography{biblio}

@ARTICLE{Ossenkopf_Henning_1994,
       author = {{Ossenkopf}, V. and {Henning}, Th.},
        title = "{Dust opacities for protostellar cores.}",
      journal = {\aap},
         year = 1994,
        month = nov,
       volume = {291},
        pages = {943-959},
       adsurl = {https://ui.adsabs.harvard.edu/abs/1994A&A...291..943O}
}

@ARTICLE{Paola_Dalessio_1999,
       author = {{D'Alessio}, Paola and {Calvet}, Nuria and {Hartmann}, Lee and {Lizano}, Susana and {Cant{\'o}}, Jorge},
        title = "{Accretion Disks around Young Objects. II. Tests of Well-mixed Models with ISM Dust}",
      journal = {\apj},
         year = 1999,
        month = dec,
       volume = {527},
       number = {2},
        pages = {893-909},
          doi = {10.1086/308103},
archivePrefix = {arXiv},
       eprint = {astro-ph/9907330},
 primaryClass = {astro-ph},
       adsurl = {https://ui.adsabs.harvard.edu/abs/1999ApJ...527..893D}
}

@ARTICLE{LyndenBell_Pringle_1974,
       author = {{Lynden-Bell}, D. and {Pringle}, J.~E.},
        title = "{The evolution of viscous discs and the origin of the nebular variables.}",
      journal = {\mnras},
         year = 1974,
        month = sep,
       volume = {168},
        pages = {603-637},
          doi = {10.1093/mnras/168.3.603},
       adsurl = {https://ui.adsabs.harvard.edu/abs/1974MNRAS.168..603L}
}

@ARTICLE{FehrAndrews,
       author = {{Fehr}, Anna J. and {Andrews}, Sean M.},
        title = "{Measuring the Two-Dimensional Thermal Structures of Protoplanetary Disks}",
      journal = {arXiv e-prints},
         year = 2025,
        month = sep,
          eid = {arXiv:2509.15196},
        pages = {arXiv:2509.15196},
          doi = {10.48550/arXiv.2509.15196},
archivePrefix = {arXiv},
       eprint = {2509.15196},
 primaryClass = {astro-ph.EP},
       adsurl = {https://ui.adsabs.harvard.edu/abs/2025arXiv250915196F}
}

@ARTICLE{Pezzotta,
       author = {{Pezzotta}, V. and {Facchini}, S. and {Longarini}, C. and {Lodato}, G. and {Martire}, P.},
        title = "{The two-dimensional pressure structure of the HD 163296 protoplanetary disk as probed by multi-molecule kinematics}",
      journal = {\aap},
         year = 2025,
        month = feb,
       volume = {694},
          eid = {A108},
        pages = {A108},
          doi = {10.1051/0004-6361/202451307},
archivePrefix = {arXiv},
       eprint = {2501.05517},
 primaryClass = {astro-ph.EP},
       adsurl = {https://ui.adsabs.harvard.edu/abs/2025A&A...694A.108P}
}

@ARTICLE{Rosotti_2020_b,
       author = {{Rosotti}, Giovanni P. and {Teague}, Richard and {Dullemond}, Cornelis and {Booth}, Richard A. and {Clarke}, Cathie J.},
        title = "{The efficiency of dust trapping in ringed protoplanetary discs}",
      journal = {\mnras},
         year = 2020,
        month = jun,
       volume = {495},
       number = {1},
        pages = {173-181},
          doi = {10.1093/mnras/staa1170},
archivePrefix = {arXiv},
       eprint = {2004.11394},
 primaryClass = {astro-ph.EP},
       adsurl = {https://ui.adsabs.harvard.edu/abs/2020MNRAS.495..173R}
}

@ARTICLE{Martire_2024,
       author = {{Martire}, P. and {Longarini}, C. and {Lodato}, G. and {Rosotti}, G.~P. and {Winter}, A. and {Facchini}, S. and {Hardiman}, C. and {Benisty}, M. and {Stadler}, J. and {Izquierdo}, A.~F. and {Testi}, Leonardo},
        title = "{Rotation curves in protoplanetary disks with thermal stratification. Physical model and observational evidence in MAPS disks}",
      journal = {\aap},
         year = 2024,
        month = jun,
       volume = {686},
          eid = {A9},
        pages = {A9},
          doi = {10.1051/0004-6361/202348546},
archivePrefix = {arXiv},
       eprint = {2402.12236},
 primaryClass = {astro-ph.EP},
       adsurl = {https://ui.adsabs.harvard.edu/abs/2024A&A...686A...9M}
}

@ARTICLE{Pinte_2018,
       author = {{Pinte}, C. and {M{\'e}nard}, F. and {Duch{\^e}ne}, G. and {Hill}, T. and {Dent}, W.~R.~F. and {Woitke}, P. and {Maret}, S. and {van der Plas}, G. and {Hales}, A. and {Kamp}, I. and et al.},
        title = "{Direct mapping of the temperature and velocity gradients in discs. Imaging the vertical CO snow line around IM Lupi}",
      journal = {\aap},
         year = 2018,
        month = jan,
       volume = {609},
          eid = {A47},
        pages = {A47},
          doi = {10.1051/0004-6361/201731377},
archivePrefix = {arXiv},
       eprint = {1710.06450},
 primaryClass = {astro-ph.SR},
       adsurl = {https://ui.adsabs.harvard.edu/abs/2018A&A...609A..47P}
}

@ARTICLE{Law_2021,
       author = {{Law}, Charles J. and {Teague}, Richard and {Loomis}, Ryan A. and {Bae}, Jaehan and {{\"O}berg}, Karin I. and {Czekala}, Ian and {Andrews}, Sean M. and {Aikawa}, Yuri and {Alarc{\'o}n}, Felipe and {Bergin}, Edwin A. and {Bergner}, Jennifer B. and {Booth}, Alice S. and {Bosman}, Arthur D. and {Calahan}, Jenny K. and {Cataldi}, Gianni and {Cleeves}, L. Ilsedore and {Furuya}, Kenji and {Guzm{\'a}n}, Viviana V. and {Huang}, Jane and {Ilee}, John D. and {Le Gal}, Romane and {Liu}, Yao and {Long}, Feng and {M{\'e}nard}, Fran{\c{c}}ois and {Nomura}, Hideko and {P{\'e}rez}, Laura M. and {Qi}, Chunhua and {Schwarz}, Kamber R. and {Soto}, Daniela and {Tsukagoshi}, Takashi and {Yamato}, Yoshihide and {van't Hoff}, Merel L.~R. and {Walsh}, Catherine and {Wilner}, David J. and {Zhang}, Ke},
        title = "{Molecules with ALMA at Planet-forming Scales (MAPS). IV. Emission Surfaces and Vertical Distribution of Molecules}",
      journal = {\apjs},
         year = 2021,
        month = nov,
       volume = {257},
       number = {1},
          eid = {4},
        pages = {4},
          doi = {10.3847/1538-4365/ac1439},
archivePrefix = {arXiv},
       eprint = {2109.06217},
 primaryClass = {astro-ph.GA},
       adsurl = {https://ui.adsabs.harvard.edu/abs/2021ApJS..257....4L}
}

@ARTICLE{Law_2022,
       author = {{Law}, Charles J. and {Crystian}, Sage and {Teague}, Richard and {{\"O}berg}, Karin I. and {Rich}, Evan A. and {Andrews}, Sean M. and {Bae}, Jaehan and {Flaherty}, Kevin and {Guzm{\'a}n}, Viviana V. and {Huang}, Jane and {Ilee}, John D. and {Kastner}, Joel H. and {Loomis}, Ryan A. and {Long}, Feng and {P{\'e}rez}, Laura M. and {P{\'e}rez}, Sebasti{\'a}n and {Qi}, Chunhua and {Rosotti}, Giovanni P. and {Ru{\'\i}z-Rodr{\'\i}guez}, Dary and {Tsukagoshi}, Takashi and {Wilner}, David J.},
        title = "{CO Line Emission Surfaces and Vertical Structure in Midinclination Protoplanetary Disks}",
      journal = {\apj},
         year = 2022,
        month = jun,
       volume = {932},
       number = {2},
          eid = {114},
        pages = {114},
          doi = {10.3847/1538-4357/ac6c02},
archivePrefix = {arXiv},
       eprint = {2205.01776},
 primaryClass = {astro-ph.EP},
       adsurl = {https://ui.adsabs.harvard.edu/abs/2022ApJ...932..114L}
}

@ARTICLE{Law_2023,
       author = {{Law}, Charles J. and {Teague}, Richard and {{\"O}berg}, Karin I. and {Rich}, Evan A. and {Andrews}, Sean M. and {Bae}, Jaehan and {Benisty}, Myriam and {Facchini}, Stefano and {Flaherty}, Kevin and {Isella}, Andrea and {Jin}, Sheng and {Hashimoto}, Jun and {Huang}, Jane and {Loomis}, Ryan A. and {Long}, Feng and {Romero-Mirza}, Carlos E. and {Paneque-Carre{\~n}o}, Teresa and {P{\'e}rez}, Laura M. and {Qi}, Chunhua and {Schwarz}, Kamber R. and {Stadler}, Jochen and {Tsukagoshi}, Takashi and {Wilner}, David J. and {van der Plas}, Gerrit},
        title = "{Mapping Protoplanetary Disk Vertical Structure with CO Isotopologue Line Emission}",
      journal = {\apj},
         year = 2023,
        month = may,
       volume = {948},
       number = {1},
          eid = {60},
        pages = {60},
          doi = {10.3847/1538-4357/acb3c4},
archivePrefix = {arXiv},
       eprint = {2212.08667},
 primaryClass = {astro-ph.EP},
       adsurl = {https://ui.adsabs.harvard.edu/abs/2023ApJ...948...60L}
}

@ARTICLE{exoALMA_temperature,
       author = {{Galloway-Sprietsma}, Maria and {Bae}, Jaehan and {Izquierdo}, Andr{\'e}s F. and {Stadler}, Jochen and {Longarini}, Cristiano and {Teague}, Richard and {Andrews}, Sean M. and {Winter}, Andrew J. and {Benisty}, Myriam and {Facchini}, Stefano and {Rosotti}, Giovanni and {Zawadzki}, Brianna and {Pinte}, Christophe and {Fasano}, Daniele and {Barraza-Alfaro}, Marcelo and {Cataldi}, Gianni and {Cuello}, Nicol{\'a}s and {Curone}, Pietro and {Czekala}, Ian and {Flock}, Mario and {Fukagawa}, Misato and {Gardner}, Charles H. and {Garg}, Himanshi and {Hall}, Cassandra and {Huang}, Jane and {Ilee}, John D. and {Kanagawa}, Kazuhiro and {Lesur}, Geoffroy and {Lodato}, Giuseppe and {Loomis}, Ryan A. and {Menard}, Francois and {Orihara}, Ryuta and {Price}, Daniel J. and {Wafflard-Fernandez}, Gaylor and {Wilner}, David J. and {W{\"o}lfer}, Lisa and {Yen}, Hsi-Wei and {Yoshida}, Tomohiro C.},
        title = "{exoALMA. V. Gaseous Emission Surfaces and Temperature Structures}",
      journal = {\apjl},
         year = 2025,
        month = may,
       volume = {984},
       number = {1},
          eid = {L10},
        pages = {L10},
          doi = {10.3847/2041-8213/adc437},
archivePrefix = {arXiv},
       eprint = {2504.19902},
 primaryClass = {astro-ph.EP},
       adsurl = {https://ui.adsabs.harvard.edu/abs/2025ApJ...984L..10G}
}

@ARTICLE{Birnstiel_2024,
       author = {{Birnstiel}, Tilman},
        title = "{Dust Growth and Evolution in Protoplanetary Disks}",
      journal = {\araa},
         year = 2024,
        month = sep,
       volume = {62},
       number = {1},
        pages = {157-202},
          doi = {10.1146/annurev-astro-071221-052705},
archivePrefix = {arXiv},
       eprint = {2312.13287},
 primaryClass = {astro-ph.EP},
       adsurl = {https://ui.adsabs.harvard.edu/abs/2024ARA&A..62..157B}
}

@ARTICLE{Henyey_Greenstein_1941,
       author = {{Henyey}, L.~G. and {Greenstein}, J.~L.},
        title = "{Diffuse radiation in the Galaxy.}",
      journal = {\apj},
         year = 1941,
        month = jan,
       volume = {93},
        pages = {70-83},
          doi = {10.1086/144246},
       adsurl = {https://ui.adsabs.harvard.edu/abs/1941ApJ....93...70H}
}

@ARTICLE{Dullemond_Dominik_2004,
       author = {{Dullemond}, C.~P. and {Dominik}, C.},
        title = "{The effect of dust settling on the appearance  of protoplanetary disks}",
      journal = {\aap},
         year = 2004,
        month = jul,
       volume = {421},
        pages = {1075-1086},
          doi = {10.1051/0004-6361:20040284},
archivePrefix = {arXiv},
       eprint = {astro-ph/0405226},
 primaryClass = {astro-ph},
       adsurl = {https://ui.adsabs.harvard.edu/abs/2004A&A...421.1075D}
}

@ARTICLE{Epstein_1924,
       author = {{Epstein}, Paul S.},
        title = "{On the Resistance Experienced by Spheres in their Motion through Gases}",
      journal = {Physical Review},
         year = 1924,
        month = jun,
       volume = {23},
       number = {6},
        pages = {710-733},
          doi = {10.1103/PhysRev.23.710},
       adsurl = {https://ui.adsabs.harvard.edu/abs/1924PhRv...23..710E}
}

@ARTICLE{Villenave_2020,
       author = {{Villenave}, M. and {M{\'e}nard}, F. and {Dent}, W.~R.~F. and {Duch{\^e}ne}, G. and {Stapelfeldt}, K.~R. and {Benisty}, M. and {Boehler}, Y. and {van der Plas}, G. and {Pinte}, C. and {Telkamp}, Z. and et al.},
        title = "{Observations of edge-on protoplanetary disks with ALMA. I. Results from continuum data}",
      journal = {\aap},
         year = 2020,
        month = oct,
       volume = {642},
          eid = {A164},
        pages = {A164},
          doi = {10.1051/0004-6361/202038087},
archivePrefix = {arXiv},
       eprint = {2008.06518},
 primaryClass = {astro-ph.SR},
       adsurl = {https://ui.adsabs.harvard.edu/abs/2020A&A...642A.164V}
}

@ARTICLE{Shakura_Sunyaev_1973,
       author = {{Shakura}, N.~I. and {Sunyaev}, R.~A.},
        title = "{Black holes in binary systems. Observational appearance.}",
      journal = {\aap},
         year = 1973,
        month = jan,
       volume = {24},
        pages = {337-355},
       adsurl = {https://ui.adsabs.harvard.edu/abs/1973A&A....24..337S}
}

@ARTICLE{Pinte_2006,
       author = {{Pinte}, C. and {M{\'e}nard}, F. and {Duch{\^e}ne}, G. and {Bastien}, P.},
        title = "{Monte Carlo radiative transfer in protoplanetary disks}",
      journal = {\aap},
         year = 2006,
        month = dec,
       volume = {459},
       number = {3},
        pages = {797-804},
          doi = {10.1051/0004-6361:20053275},
archivePrefix = {arXiv},
       eprint = {astro-ph/0606550},
 primaryClass = {astro-ph},
       adsurl = {https://ui.adsabs.harvard.edu/abs/2006A&A...459..797P}
}

@ARTICLE{Pinte_2009,
       author = {{Pinte}, C. and {Harries}, T.~J. and {Min}, M. and {Watson}, A.~M. and {Dullemond}, C.~P. and {Woitke}, P. and {M{\'e}nard}, F. and {Dur{\'a}n-Rojas}, M.~C.},
        title = "{Benchmark problems for continuum radiative transfer. High optical depths, anisotropic scattering, and polarisation}",
      journal = {\aap},
         year = 2009,
        month = may,
       volume = {498},
       number = {3},
        pages = {967-980},
          doi = {10.1051/0004-6361/200811555},
archivePrefix = {arXiv},
       eprint = {0903.1231},
 primaryClass = {astro-ph.SR},
       adsurl = {https://ui.adsabs.harvard.edu/abs/2009A&A...498..967P}
}

@ARTICLE{Draine_Lee_1984,
       author = {{Draine}, B.~T. and {Lee}, H.~M.},
        title = "{Optical Properties of Interstellar Graphite and Silicate Grains}",
      journal = {\apj},
         year = 1984,
        month = oct,
       volume = {285},
        pages = {89},
          doi = {10.1086/162480},
       adsurl = {https://ui.adsabs.harvard.edu/abs/1984ApJ...285...89D}
}

@INPROCEEDINGS{Mathis_1997,
       author = {{Mathis}, John S.},
        title = "{Composition and Size of Interstellar Dust}",
    booktitle = {From Stardust to Planetesimals},
         year = 1997,
       editor = {{Pendleton}, Yvonne J.},
       series = {Astronomical Society of the Pacific Conference Series},
       volume = {122},
        month = jan,
        pages = {87},
       adsurl = {https://ui.adsabs.harvard.edu/abs/1997ASPC..122...87M}
}

@ARTICLE{Avenhaus2018,
       author = {{Avenhaus}, Henning and {Quanz}, Sascha P. and {Garufi}, Antonio and {Perez}, Sebastian and {Casassus}, Simon and {Pinte}, Christophe and {Bertrang}, Gesa H. -M. and {Caceres}, Claudio and {Benisty}, Myriam and {Dominik}, Carsten},
        title = "{Disks around T Tauri Stars with SPHERE (DARTTS-S). I. SPHERE/IRDIS Polarimetric Imaging of Eight Prominent T Tauri Disks}",
      journal = {\apj},
         year = 2018,
        month = aug,
       volume = {863},
       number = {1},
          eid = {44},
        pages = {44},
          doi = {10.3847/1538-4357/aab846},
archivePrefix = {arXiv},
       eprint = {1803.10882},
 primaryClass = {astro-ph.SR},
       adsurl = {https://ui.adsabs.harvard.edu/abs/2018ApJ...863...44A}
}

@ARTICLE{Ginski2023,
       author = {{Ginski}, Christian and {Tazaki}, Ryo and {Dominik}, Carsten and {Stolker}, Tomas},
        title = "{Observed Polarized Scattered Light Phase Functions of Planet-forming Disks}",
      journal = {\apj},
         year = 2023,
        month = aug,
       volume = {953},
       number = {1},
          eid = {92},
        pages = {92},
          doi = {10.3847/1538-4357/acdc97},
archivePrefix = {arXiv},
       eprint = {2301.04617},
 primaryClass = {astro-ph.EP},
       adsurl = {https://ui.adsabs.harvard.edu/abs/2023ApJ...953...92G}
}

@ARTICLE{DeBoer2021,
       author = {{de Boer}, J. and {Ginski}, C. and {Chauvin}, G. and {M{\'e}nard}, F. and {Benisty}, M. and {Dominik}, C. and {Maaskant}, K. and {Girard}, J.~H. and {van der Plas}, G. and {Garufi}, A. and {Perrot}, C. and {Stolker}, T. and {Avenhaus}, H. and {Bohn}, A. and {Delboulb{\'e}}, A. and {Jaquet}, M. and {Buey}, T. and {M{\"o}ller-Nilsson}, O. and {Pragt}, J. and {Fusco}, T.},
        title = "{Possible single-armed spiral in the protoplanetary disk around HD 34282}",
      journal = {\aap},
         year = 2021,
        month = may,
       volume = {649},
          eid = {A25},
        pages = {A25},
          doi = {10.1051/0004-6361/201936787},
archivePrefix = {arXiv},
       eprint = {2010.12202},
 primaryClass = {astro-ph.EP},
       adsurl = {https://ui.adsabs.harvard.edu/abs/2021A&A...649A..25D}
}

@ARTICLE{Ginski_2016,
       author = {{Ginski}, C. and {Stolker}, T. and {Pinilla}, P. and {Dominik}, C. and {Boccaletti}, A. and {de Boer}, J. and {Benisty}, M. and {Biller}, B. and {Feldt}, M. and {Garufi}, A. and {Keller}, C.~U. and {Kenworthy}, M. and {Maire}, A.~L. and {M{\'e}nard}, F. and {Mesa}, D. and {Milli}, J. and {Min}, M. and {Pinte}, C. and {Quanz}, S.~P. and {van Boekel}, R. and {Bonnefoy}, M. and {Chauvin}, G. and {Desidera}, S. and {Gratton}, R. and {Girard}, J.~H.~V. and {Keppler}, M. and {Kopytova}, T. and {Lagrange}, A. -M. and {Langlois}, M. and {Rouan}, D. and {Vigan}, A.},
        title = "{Direct detection of scattered light gaps in the transitional disk around HD 97048 with VLT/SPHERE}",
      journal = {\aap},
         year = 2016,
        month = nov,
       volume = {595},
          eid = {A112},
        pages = {A112},
          doi = {10.1051/0004-6361/201629265},
archivePrefix = {arXiv},
       eprint = {1609.04027},
 primaryClass = {astro-ph.EP},
       adsurl = {https://ui.adsabs.harvard.edu/abs/2016A&A...595A.112G}
}

@ARTICLE{Pollack_1994,
       author = {{Pollack}, James B. and {Hollenbach}, David and {Beckwith}, Steven and {Simonelli}, Damon P. and {Roush}, Ted and {Fong}, Wesley},
        title = "{Composition and Radiative Properties of Grains in Molecular Clouds and Accretion Disks}",
      journal = {\apj},
         year = 1994,
        month = feb,
       volume = {421},
        pages = {615},
          doi = {10.1086/173677},
       adsurl = {https://ui.adsabs.harvard.edu/abs/1994ApJ...421..615P}
}

@ARTICLE{Min_2016,
       author = {{Min}, M. and {Rab}, Ch. and {Woitke}, P. and {Dominik}, C. and {M{\'e}nard}, F.},
        title = "{Multiwavelength optical properties of compact dust aggregates in protoplanetary disks}",
      journal = {\aap},
         year = 2016,
        month = jan,
       volume = {585},
          eid = {A13},
        pages = {A13},
          doi = {10.1051/0004-6361/201526048},
archivePrefix = {arXiv},
       eprint = {1510.05426},
 primaryClass = {astro-ph.EP},
       adsurl = {https://ui.adsabs.harvard.edu/abs/2016A&A...585A..13M}
}

@ARTICLE{Kataoka_2015,
       author = {{Kataoka}, Akimasa and {Muto}, Takayuki and {Momose}, Munetake and {Tsukagoshi}, Takashi and {Fukagawa}, Misato and {Shibai}, Hiroshi and {Hanawa}, Tomoyuki and {Murakawa}, Koji and {Dullemond}, Cornelis P.},
        title = "{Millimeter-wave Polarization of Protoplanetary Disks due to Dust Scattering}",
      journal = {\apj},
         year = 2015,
        month = aug,
       volume = {809},
       number = {1},
          eid = {78},
        pages = {78},
          doi = {10.1088/0004-637X/809/1/78},
archivePrefix = {arXiv},
       eprint = {1504.04812},
 primaryClass = {astro-ph.EP},
       adsurl = {https://ui.adsabs.harvard.edu/abs/2015ApJ...809...78K}
}

@ARTICLE{Mass_HD163296,
       author = {{van den Ancker}, M.~E. and {Bouwman}, J. and {Wesselius}, P.~R. and {Waters}, L.~B.~F.~M. and {Dougherty}, S.~M. and {van Dishoeck}, E.~F.},
        title = "{ISO spectroscopy of circumstellar dust in the Herbig Ae systems AB Aur and HD 163296}",
      journal = {\aap},
         year = 2000,
        month = may,
       volume = {357},
        pages = {325-329},
          doi = {10.48550/arXiv.astro-ph/0002440},
archivePrefix = {arXiv},
       eprint = {astro-ph/0002440},
 primaryClass = {astro-ph},
       adsurl = {https://ui.adsabs.harvard.edu/abs/2000A&A...357..325V}
}

@ARTICLE{Flaherty2020,
       author = {{Flaherty}, Kevin and {Hughes}, A. Meredith and {Simon}, Jacob B. and {Qi}, Chunhua and {Bai}, Xue-Ning and {Bulatek}, Alyssa and {Andrews}, Sean M. and {Wilner}, David J. and {K{\'o}sp{\'a}l}, {\'A}gnes},
        title = "{Measuring Turbulent Motion in Planet-forming Disks with ALMA: A Detection around DM Tau and Nondetections around MWC 480 and V4046 Sgr}",
      journal = {\apj},
         year = 2020,
        month = jun,
       volume = {895},
       number = {2},
          eid = {109},
        pages = {109},
          doi = {10.3847/1538-4357/ab8cc5},
archivePrefix = {arXiv},
       eprint = {2004.12176},
 primaryClass = {astro-ph.SR},
       adsurl = {https://ui.adsabs.harvard.edu/abs/2020ApJ...895..109F}
}

@ARTICLE{Villenave2019,
       author = {{Villenave}, M. and {Benisty}, M. and {Dent}, W.~R.~F. and {M{\'e}nard}, F. and {Garufi}, A. and {Ginski}, C. and {Pinilla}, P. and {Pinte}, C. and {Williams}, J.~P. and {de Boer}, J. and {Morino}, J. -I. and {Fukagawa}, M. and {Dominik}, C. and {Flock}, M. and {Henning}, T. and {Juh{\'a}sz}, A. and {Keppler}, M. and {Muro-Arena}, G. and {Olofsson}, J. and {P{\'e}rez}, L.~M. and {van der Plas}, G. and {Zurlo}, A. and {Carle}, M. and {Feautrier}, P. and {Pavlov}, A. and {Pragt}, J. and {Ramos}, J. and {Sauvage}, J. -F. and {Stadler}, E. and {Weber}, L.},
        title = "{Spatial segregation of dust grains in transition disks. SPHERE observations of 2MASS J16083070-3828268 and RXJ1852.3-3700}",
      journal = {\aap},
         year = 2019,
        month = apr,
       volume = {624},
          eid = {A7},
        pages = {A7},
          doi = {10.1051/0004-6361/201834800},
archivePrefix = {arXiv},
       eprint = {1902.04612},
 primaryClass = {astro-ph.SR},
       adsurl = {https://ui.adsabs.harvard.edu/abs/2019A&A...624A...7V}
}

@ARTICLE{Ginski_2024,
       author = {{Ginski}, C. and {Garufi}, A. and {Benisty}, M. and {Tazaki}, R. and {Dominik}, C. and {Ribas}, {\'A}. and {Engler}, N. and {Birnstiel}, T. and {Chauvin}, G. and {Columba}, G. and et al.},
        title = "{The SPHERE view of the Chamaeleon I star-forming region. The full census of planet-forming disks with GTO and DESTINYS programs}",
      journal = {\aap},
         year = 2024,
        month = may,
       volume = {685},
          eid = {A52},
        pages = {A52},
          doi = {10.1051/0004-6361/202244005},
archivePrefix = {arXiv},
       eprint = {2403.02149},
 primaryClass = {astro-ph.GA},
       adsurl = {https://ui.adsabs.harvard.edu/abs/2024A&A...685A..52G}
}

@ARTICLE{Maps_Teague,
       author = {{Teague}, Richard and {Bae}, Jaehan and {Aikawa}, Yuri and {Andrews}, Sean M. and {Bergin}, Edwin A. and {Bergner}, Jennifer B. and {Boehler}, Yann and {Booth}, Alice S. and {Bosman}, Arthur D. and {Cataldi}, Gianni and {Czekala}, Ian and {Guzm{\'a}n}, Viviana V. and {Huang}, Jane and {Ilee}, John D. and {Law}, Charles J. and {Le Gal}, Romane and {Long}, Feng and {Loomis}, Ryan A. and {M{\'e}nard}, Fran{\c{c}}ois and {{\"O}berg}, Karin I. and {P{\'e}rez}, Laura M. and {Schwarz}, Kamber R. and {Sierra}, Anibal and {Walsh}, Catherine and {Wilner}, David J. and {Yamato}, Yoshihide and {Zhang}, Ke},
        title = "{Molecules with ALMA at Planet-forming Scales (MAPS). XVIII. Kinematic Substructures in the Disks of HD 163296 and MWC 480}",
      journal = {\apjs},
         year = 2021,
        month = nov,
       volume = {257},
       number = {1},
          eid = {18},
        pages = {18},
          doi = {10.3847/1538-4365/ac1438},
archivePrefix = {arXiv},
       eprint = {2109.06218},
 primaryClass = {astro-ph.EP},
       adsurl = {https://ui.adsabs.harvard.edu/abs/2021ApJS..257...18T}
}

@ARTICLE{exoALMA_rotation,
       author = {{Longarini}, Cristiano and {Lodato}, Giuseppe and {Rosotti}, Giovanni and {Andrews}, Sean and {Winter}, Andrew and {Stadler}, Jochen and {Izquierdo}, Andr{\'e}s and {Galloway-Sprietsma}, Maria and {Facchini}, Stefano and {Curone}, Pietro and {Benisty}, Myriam and {Teague}, Richard and {Bae}, Jaehan and {Barraza-Alfaro}, Marcelo and {Cataldi}, Gianni and {Czekala}, Ian and {Cuello}, Nicol{\'a}s and {Fasano}, Daniele and {Flock}, Mario and {Fukagawa}, Misato and {Garg}, Himanshi and {Hall}, Cassandra and {Hammond}, Iain and {Hardiman}, Caitlyn and {Hilder}, Thomas and {Huang}, Jane and {Ilee}, John D. and {Isella}, Andrea and {Kanagawa}, Kazuhiro and {Lesur}, Geoffroy and {Loomis}, Ryan A. and {M{\'e}nard}, Francois and {Orihara}, Ryuta and {Pinte}, Christophe and {Price}, Daniel and {Testi}, Leonardo and {Fernandez}, Gaylor Wafflard- and {W{\"o}lfer}, Lisa and {Yen}, Hsi-Wei and {Yoshida}, Tomohiro C. and {Zawadzki}, Brianna},
        title = "{exoALMA. XII. Weighing and Sizing exoALMA Disks with Rotation Curve Modelling}",
      journal = {\apjl},
         year = 2025,
        month = may,
       volume = {984},
       number = {1},
          eid = {L17},
        pages = {L17},
          doi = {10.3847/2041-8213/adc431},
       adsurl = {https://ui.adsabs.harvard.edu/abs/2025ApJ...984L..17L}
}

@ARTICLE{Sierra_2021,
       author = {{Sierra}, Anibal and {P{\'e}rez}, Laura M. and {Zhang}, Ke and {Law}, Charles J. and {Guzm{\'a}n}, Viviana V. and {Qi}, Chunhua and {Bosman}, Arthur D. and {{\"O}berg}, Karin I. and {Andrews}, Sean M. and {Long}, Feng and et al.},
        title = "{Molecules with ALMA at Planet-forming Scales (MAPS). XIV. Revealing Disk Substructures in Multiwavelength Continuum Emission}",
      journal = {\apjs},
         year = 2021,
        month = nov,
       volume = {257},
       number = {1},
          eid = {14},
        pages = {14},
          doi = {10.3847/1538-4365/ac1431},
archivePrefix = {arXiv},
       eprint = {2109.06433},
 primaryClass = {astro-ph.EP},
       adsurl = {https://ui.adsabs.harvard.edu/abs/2021ApJS..257...14S}
}

@ARTICLE{Tanaka_1996,
       author = {{Tanaka}, Hidekazu and {Inaba}, Satoshi and {Nakazawa}, Kiyoshi},
        title = "{Steady-State Size Distribution for the Self-Similar Collision Cascade}",
      journal = {\icarus},
         year = 1996,
        month = oct,
       volume = {123},
       number = {2},
        pages = {450-455},
          doi = {10.1006/icar.1996.0170},
       adsurl = {https://ui.adsabs.harvard.edu/abs/1996Icar..123..450T}
}

@ARTICLE{Williams_1994,
       author = {{Williams}, D.~R. and {Wetherill}, G.~W.},
        title = "{Size Distribution of Collisionally Evolved Asteroidal Populations: Analytical Solution for Self-Similar Collision Cascades}",
      journal = {\icarus},
         year = 1994,
        month = jan,
       volume = {107},
       number = {1},
        pages = {117-128},
          doi = {10.1006/icar.1994.1010},
       adsurl = {https://ui.adsabs.harvard.edu/abs/1994Icar..107..117W}
}

@ARTICLE{Birnstiel_2012,
       author = {{Birnstiel}, T. and {Klahr}, H. and {Ercolano}, B.},
        title = "{A simple model for the evolution of the dust population in protoplanetary disks}",
      journal = {\aap},
         year = 2012,
        month = mar,
       volume = {539},
          eid = {A148},
        pages = {A148},
          doi = {10.1051/0004-6361/201118136},
archivePrefix = {arXiv},
       eprint = {1201.5781},
 primaryClass = {astro-ph.EP},
       adsurl = {https://ui.adsabs.harvard.edu/abs/2012A&A...539A.148B}
}

@ARTICLE{Birnstiel_2015,
       author = {{Birnstiel}, Tilman and {Andrews}, Sean M. and {Pinilla}, Paola and {Kama}, Mihkel},
        title = "{Dust Evolution Can Produce Scattered Light Gaps in Protoplanetary Disks}",
      journal = {\apjl},
         year = 2015,
        month = nov,
       volume = {813},
       number = {1},
          eid = {L14},
        pages = {L14},
          doi = {10.1088/2041-8205/813/1/L14},
archivePrefix = {arXiv},
       eprint = {1510.05660},
 primaryClass = {astro-ph.EP},
       adsurl = {https://ui.adsabs.harvard.edu/abs/2015ApJ...813L..14B}
}

@ARTICLE{Franceschi_2023,
       author = {{Franceschi}, Riccardo and {Birnstiel}, Tilman and {Henning}, Thomas and {Sharma}, Anirudh},
        title = "{Constraining the turbulence and the dust disk in IM Lup: Onset of planetesimal formation}",
      journal = {\aap},
         year = 2023,
        month = mar,
       volume = {671},
          eid = {A125},
        pages = {A125},
          doi = {10.1051/0004-6361/202244869},
archivePrefix = {arXiv},
       eprint = {2212.01291},
 primaryClass = {astro-ph.EP},
       adsurl = {https://ui.adsabs.harvard.edu/abs/2023A&A...671A.125F}
}

@ARTICLE{Oberg_2023,
       author = {{{\"O}berg}, Karin I. and {Facchini}, Stefano and {Anderson}, Dana E.},
        title = "{Protoplanetary Disk Chemistry}",
      journal = {\araa},
         year = 2023,
        month = aug,
       volume = {61},
        pages = {287-328},
          doi = {10.1146/annurev-astro-022823-040820},
archivePrefix = {arXiv},
       eprint = {2309.05685},
 primaryClass = {astro-ph.EP},
       adsurl = {https://ui.adsabs.harvard.edu/abs/2023ARA&A..61..287O}
}

@software{Dominik_2021,
       author = {{Dominik}, Carsten and {Min}, Michiel and {Tazaki}, Ryo},
        title = "{OpTool: Command-line driven tool for creating complex dust opacities}",
 howpublished = {Astrophysics Source Code Library, record ascl:2104.010},
         year = 2021,
        month = apr,
          eid = {ascl:2104.010},
archivePrefix = {ascl},
       eprint = {2104.010},
       adsurl = {https://ui.adsabs.harvard.edu/abs/2021ascl.soft04010D}
}

@ARTICLE{Walsh_2016,
       author = {{Walsh}, Catherine and {Juh{\'a}sz}, Attila and {Meeus}, Gwendolyn and {Dent}, William R.~F. and {Maud}, Luke T. and {Aikawa}, Yuri and {Millar}, Tom J. and {Nomura}, Hideko},
        title = "{ALMA Reveals the Anatomy of the mm-sized Dust and Molecular Gas in the HD 97048 Disk}",
      journal = {\apj},
         year = 2016,
        month = nov,
       volume = {831},
       number = {2},
          eid = {200},
        pages = {200},
          doi = {10.3847/0004-637X/831/2/200},
archivePrefix = {arXiv},
       eprint = {1609.02011},
 primaryClass = {astro-ph.EP},
       adsurl = {https://ui.adsabs.harvard.edu/abs/2016ApJ...831..200W}
}

@ARTICLE{Dullemond_2001,
       author = {{Dullemond}, C.~P. and {Dominik}, C. and {Natta}, A.},
        title = "{Passive Irradiated Circumstellar Disks with an Inner Hole}",
      journal = {\apj},
         year = 2001,
        month = oct,
       volume = {560},
       number = {2},
        pages = {957-969},
          doi = {10.1086/323057},
archivePrefix = {arXiv},
       eprint = {astro-ph/0106470},
 primaryClass = {astro-ph},
       adsurl = {https://ui.adsabs.harvard.edu/abs/2001ApJ...560..957D}
}

@ARTICLE{Curone_2025,
       author = {{Curone}, Pietro and {Facchini}, Stefano and {Andrews}, Sean M. and {Testi}, Leonardo and {Benisty}, Myriam and {Czekala}, Ian and {Huang}, Jane and {Ilee}, John D. and {Isella}, Andrea and {Lodato}, Giuseppe and {Loomis}, Ryan A. and {Stadler}, Jochen and {Winter}, Andrew J. and {Bae}, Jaehan and {Barraza-Alfaro}, Marcelo and {Cataldi}, Gianni and {Cuello}, Nicol{\'a}s and {Fasano}, Daniele and {Flock}, Mario and {Fukagawa}, Misato and {Galloway-Sprietsma}, Maria and {Garg}, Himanshi and {Hall}, Cassandra and {Izquierdo}, Andr{\'e}s F. and {Kanagawa}, Kazuhiro and {Lesur}, Geoffroy and {Longarini}, Cristiano and {Menard}, Francois and {Orihara}, Ryuta and {Pinte}, Christophe and {Price}, Daniel J. and {Rosotti}, Giovanni and {Teague}, Richard and {Wafflard-Fernandez}, Gaylor and {Wilner}, David J. and {W{\"o}lfer}, Lisa and {Yen}, Hsi-Wei and {Yoshida}, Tomohiro C. and {Zawadzki}, Brianna},
        title = "{exoALMA. IV. Substructures, Asymmetries, and the Faint Outer Disk in Continuum Emission}",
      journal = {\apjl},
         year = 2025,
        month = may,
       volume = {984},
       number = {1},
          eid = {L9},
        pages = {L9},
          doi = {10.3847/2041-8213/adc438},
archivePrefix = {arXiv},
       eprint = {2504.18725},
 primaryClass = {astro-ph.EP},
       adsurl = {https://ui.adsabs.harvard.edu/abs/2025ApJ...984L...9C}
}

@ARTICLE{Muro-Arena_2018,
       author = {{Muro-Arena}, G.~A. and {Dominik}, C. and {Waters}, L.~B.~F.~M. and {Min}, M. and {Klarmann}, L. and {Ginski}, C. and {Isella}, A. and {Benisty}, M. and {Pohl}, A. and {Garufi}, A. and {Hagelberg}, J. and {Langlois}, M. and {Menard}, F. and {Pinte}, C. and {Sezestre}, E. and {van der Plas}, G. and {Villenave}, M. and {Delboulb{\'e}}, A. and {Magnard}, Y. and {M{\"o}ller-Nilsson}, O. and {Pragt}, J. and {Rabou}, P. and {Roelfsema}, R.},
        title = "{Dust modeling of the combined ALMA and SPHERE datasets of HD 163296. Is HD 163296 really a Meeus group II disk?}",
      journal = {\aap},
         year = 2018,
        month = jun,
       volume = {614},
          eid = {A24},
        pages = {A24},
          doi = {10.1051/0004-6361/201732299},
archivePrefix = {arXiv},
       eprint = {1802.03328},
 primaryClass = {astro-ph.EP},
       adsurl = {https://ui.adsabs.harvard.edu/abs/2018A&A...614A..24M}
}

@ARTICLE{Min_2009,
       author = {{Min}, M. and {Dullemond}, C.~P. and {Dominik}, C. and {de Koter}, A. and {Hovenier}, J.~W.},
        title = "{Radiative transfer in very optically thick circumstellar disks}",
      journal = {\aap},
         year = 2009,
        month = apr,
       volume = {497},
       number = {1},
        pages = {155-166},
          doi = {10.1051/0004-6361/200811470},
archivePrefix = {arXiv},
       eprint = {0902.3092},
 primaryClass = {astro-ph.IM},
       adsurl = {https://ui.adsabs.harvard.edu/abs/2009A&A...497..155M}
}

@ARTICLE{Rich_2021,
       author = {{Rich}, Evan A. and {Teague}, Richard and {Monnier}, John D. and {Davies}, Claire L. and {Bosman}, Arthur and {Harries}, Tim J. and {Calvet}, Nuria and {Adams}, Fred C. and {Wilner}, David and {Zhu}, Zhaohuan},
        title = "{Investigating the Relative Gas and Small Dust Grain Surface Heights in Protoplanetary Disks}",
      journal = {\apj},
         year = 2021,
        month = jun,
       volume = {913},
       number = {2},
          eid = {138},
        pages = {138},
          doi = {10.3847/1538-4357/abf92e},
archivePrefix = {arXiv},
       eprint = {2104.07821},
 primaryClass = {astro-ph.EP},
       adsurl = {https://ui.adsabs.harvard.edu/abs/2021ApJ...913..138R}
}

@ARTICLE{Tazaki_2019,
       author = {{Tazaki}, Ryo and {Tanaka}, H. and {Muto}, T. and {Kataoka}, A. and {Okuzumi}, S.},
        title = "{Effect of dust size and structure on scattered-light images of protoplanetary discs}",
      journal = {\mnras},
         year = 2019,
        month = jun,
       volume = {485},
       number = {4},
        pages = {4951-4966},
          doi = {10.1093/mnras/stz662},
archivePrefix = {arXiv},
       eprint = {1903.01890},
 primaryClass = {astro-ph.EP},
       adsurl = {https://ui.adsabs.harvard.edu/abs/2019MNRAS.485.4951T}
}

@ARTICLE{Tazaki_2023,
       author = {{Tazaki}, Ryo and {Ginski}, Christian and {Dominik}, Carsten},
        title = "{Fractal Aggregates of Submicron-sized Grains in the Young Planet-forming Disk around IM Lup}",
      journal = {\apjl},
         year = 2023,
        month = feb,
       volume = {944},
       number = {2},
          eid = {L43},
        pages = {L43},
          doi = {10.3847/2041-8213/acb824},
archivePrefix = {arXiv},
       eprint = {2302.01119},
 primaryClass = {astro-ph.EP},
       adsurl = {https://ui.adsabs.harvard.edu/abs/2023ApJ...944L..43T}
}

@ARTICLE{Columba_2025,
       author = {{Columba}, G. and {Rigliaco}, E. and {Gratton}, R. and {Ginski}, C. and {Garufi}, A. and {Benisty}, M. and {Facchini}, S. and {van Holstein}, R.~G. and {Ribas}, {\'A}. and {Williams}, J. and et al.},
        title = "{Disk Evolution Study Through Imaging of Nearby Young Stars (DESTINYS): V721 CrA and BN CrA have wide and structured disks in the polarised infrared}",
      journal = {\aap},
         year = 2026,
        month = jan,
       volume = {706},
          eid = {A16},
        pages = {A16},
          doi = {10.1051/0004-6361/202556331},
archivePrefix = {arXiv},
       eprint = {2511.01717},
 primaryClass = {astro-ph.SR},
       adsurl = {https://ui.adsabs.harvard.edu/abs/2026A&A...706A..16C}
}

@ARTICLE{Roumesy_2025,
       author = {{Roumesy}, M. and {M{\'e}nard}, F. and {Tazaki}, R. and {Duch{\^e}ne}, G. and {Martinien}, L. and {Zerna}, R.},
        title = "{DRAGyS {\textendash} A comprehensive tool for extracting scattering phase functions in protoplanetary disks: Disk ring adjusted geometry yields scattering phase function}",
      journal = {\aap},
         year = 2025,
        month = jul,
       volume = {699},
          eid = {A162},
        pages = {A162},
          doi = {10.1051/0004-6361/202554058},
archivePrefix = {arXiv},
       eprint = {2505.20070},
 primaryClass = {astro-ph.IM},
       adsurl = {https://ui.adsabs.harvard.edu/abs/2025A&A...699A.162R}
}

@INPROCEEDINGS{Benisty_2023,
       author = {{Benisty}, M. and {Dominik}, C. and {Follette}, K. and {Garufi}, A. and {Ginski}, C. and {Hashimoto}, J. and {Keppler}, M. and {Kley}, W. and {Monnier}, J.},
        title = "{Optical and Near-infrared View of Planet-forming Disks and Protoplanets}",
    booktitle = {Protostars and Planets VII},
         year = 2023,
       editor = {{Inutsuka}, S. and {Aikawa}, Y. and {Muto}, T. and {Tomida}, K. and {Tamura}, M.},
       series = {Astronomical Society of the Pacific Conference Series},
       volume = {534},
        month = jul,
        pages = {605},
          doi = {10.48550/arXiv.2203.09991},
archivePrefix = {arXiv},
       eprint = {2203.09991},
 primaryClass = {astro-ph.EP},
       adsurl = {https://ui.adsabs.harvard.edu/abs/2023ASPC..534..605B}
}

@ARTICLE{Garufi_2014,
       author = {{Garufi}, A. and {Quanz}, S.~P. and {Schmid}, H.~M. and {Avenhaus}, H. and {Buenzli}, E. and {Wolf}, S.},
        title = "{Shadows and cavities in protoplanetary disks: HD 163296, HD 141569A, and HD 150193A in polarized light}",
      journal = {\aap},
         year = 2014,
        month = aug,
       volume = {568},
          eid = {A40},
        pages = {A40},
          doi = {10.1051/0004-6361/201424262},
archivePrefix = {arXiv},
       eprint = {1406.7387},
 primaryClass = {astro-ph.SR},
       adsurl = {https://ui.adsabs.harvard.edu/abs/2014A&A...568A..40G}
}

@ARTICLE{Hashimoto_2011,
       author = {{Hashimoto}, J. and {Tamura}, M. and {Muto}, T. and {Kudo}, T. and {Fukagawa}, M. and {Fukue}, T. and {Goto}, M. and {Grady}, C.~A. and {Henning}, T. and {Hodapp}, K. and {Honda}, M. and {Inutsuka}, S. and {Kokubo}, E. and {Knapp}, G. and {McElwain}, M.~W. and {Momose}, M. and {Ohashi}, N. and {Okamoto}, Y.~K. and {Takami}, M. and {Turner}, E.~L. and {Wisniewski}, J. and {Janson}, M. and {Abe}, L. and {Brandner}, W. and {Carson}, J. and {Egner}, S. and {Feldt}, M. and {Golota}, T. and {Guyon}, O. and {Hayano}, Y. and {Hayashi}, M. and {Hayashi}, S. and {Ishii}, M. and {Kandori}, R. and {Kusakabe}, N. and {Matsuo}, T. and {Mayama}, S. and {Miyama}, S. and {Morino}, J.-I. and {Moro-Martin}, A. and {Nishimura}, T. and {Pyo}, T.-S. and {Suto}, H. and {Suzuki}, R. and {Takato}, N. and {Terada}, H. and {Thalmann}, C. and {Tomono}, D. and {Watanabe}, M. and {Yamada}, T. and {Takami}, H. and {Usuda}, T.},
        title = "{Direct Imaging of Fine Structures in Giant Planet-forming Regions of the Protoplanetary Disk Around AB Aurigae}",
      journal = {\apjl},
         year = 2011,
        month = mar,
       volume = {729},
       number = {2},
          eid = {L17},
        pages = {L17},
          doi = {10.1088/2041-8205/729/2/L17},
archivePrefix = {arXiv},
       eprint = {1102.4408},
 primaryClass = {astro-ph.SR},
       adsurl = {https://ui.adsabs.harvard.edu/abs/2011ApJ...729L..17H}
}

@ARTICLE{Tamura_2016,
       author = {{Tamura}, Motohide},
        title = "{SEEDS - Strategic explorations of exoplanets and disks with the Subaru Telescope -}",
      journal = {Proceedings of the Japan Academy, Series B},
         year = 2016,
        month = feb,
       volume = {92},
        pages = {45-55},
          doi = {10.2183/pjab.92.45},
       adsurl = {https://ui.adsabs.harvard.edu/abs/2016PJAB...92...45T}
}

@ARTICLE{Rich_2022,
       author = {{Rich}, Evan A. and {Monnier}, John D. and {Aarnio}, Alicia and {Laws}, Anna S.~E. and {Setterholm}, Benjamin R. and {Wilner}, David J. and {Calvet}, Nuria and {Harries}, Tim and {Miller}, Chris and {Davies}, Claire L. and {Adams}, Fred C. and {Andrews}, Sean M. and {Bae}, Jaehan and {Espaillat}, Catherine and {Greenbaum}, Alexandra Z. and {Hinkley}, Sasha and {Kraus}, Stefan and {Hartmann}, Lee and {Isella}, Andrea and {McClure}, Melissa and {Oppenheimer}, Rebecca and {P{\'e}rez}, Laura M. and {Zhu}, Zhaohuan},
        title = "{Gemini-LIGHTS: Herbig Ae/Be and Massive T Tauri Protoplanetary Disks Imaged with Gemini Planet Imager}",
      journal = {\aj},
         year = 2022,
        month = sep,
       volume = {164},
       number = {3},
          eid = {109},
        pages = {109},
          doi = {10.3847/1538-3881/ac7be4},
archivePrefix = {arXiv},
       eprint = {2206.05815},
 primaryClass = {astro-ph.EP},
       adsurl = {https://ui.adsabs.harvard.edu/abs/2022AJ....164..109R}
}

@ARTICLE{Ginski_2021,
       author = {{Ginski}, Christian and {Facchini}, Stefano and {Huang}, Jane and {Benisty}, Myriam and {Vaendel}, Dennis and {Stapper}, Lucas and {Dominik}, Carsten and {Bae}, Jaehan and {M{\'e}nard}, Fran{\c{c}}ois and {Muro-Arena}, Gabriela and {Hogerheijde}, Michiel R. and {McClure}, Melissa and {van Holstein}, Rob G. and {Birnstiel}, Tilman and {Boehler}, Yann and {Bohn}, Alexander and {Flock}, Mario and {Mamajek}, Eric E. and {Manara}, Carlo F. and {Pinilla}, Paola and {Pinte}, Christophe and {Ribas}, {\'A}lvaro},
        title = "{Disk Evolution Study Through Imaging of Nearby Young Stars (DESTINYS): Late Infall Causing Disk Misalignment and Dynamic Structures in SU Aur}",
      journal = {\apjl},
         year = 2021,
        month = feb,
       volume = {908},
       number = {2},
          eid = {L25},
        pages = {L25},
          doi = {10.3847/2041-8213/abdf57},
archivePrefix = {arXiv},
       eprint = {2102.08781},
 primaryClass = {astro-ph.EP},
       adsurl = {https://ui.adsabs.harvard.edu/abs/2021ApJ...908L..25G}
}

@ARTICLE{deBoer_2016,
       author = {{de Boer}, J. and {Salter}, G. and {Benisty}, M. and {Vigan}, A. and {Boccaletti}, A. and {Pinilla}, P. and {Ginski}, C. and {Juhasz}, A. and {Maire}, A.-L. and {Messina}, S. and {Desidera}, S. and {Cheetham}, A. and {Girard}, J.~H. and {Wahhaj}, Z. and {Langlois}, M. and {Bonnefoy}, M. and {Beuzit}, J.-L. and {Buenzli}, E. and {Chauvin}, G. and {Dominik}, C. and {Feldt}, M. and {Gratton}, R. and {Hagelberg}, J. and {Isella}, A. and {Janson}, M. and {Keller}, C.~U. and {Lagrange}, A.-M. and {Lannier}, J. and {Menard}, F. and {Mesa}, D. and {Mouillet}, D. and {Mugrauer}, M. and {Peretti}, S. and {Perrot}, C. and {Sissa}, E. and {Snik}, F. and {Vogt}, N. and {Zurlo}, A. and {SPHERE Consortium}},
        title = "{Multiple rings in the transition disk and companion candidates around RX J1615.3-3255. High contrast imaging with VLT/SPHERE}",
      journal = {\aap},
         year = 2016,
        month = nov,
       volume = {595},
          eid = {A114},
        pages = {A114},
          doi = {10.1051/0004-6361/201629267},
archivePrefix = {arXiv},
       eprint = {1610.04038},
 primaryClass = {astro-ph.EP},
       adsurl = {https://ui.adsabs.harvard.edu/abs/2016A&A...595A.114D}
}

@ARTICLE{Chiang_Goldreich_1997,
       author = {{Chiang}, E.~I. and {Goldreich}, P.},
        title = "{Spectral Energy Distributions of T Tauri Stars with Passive Circumstellar Disks}",
      journal = {\apj},
         year = 1997,
        month = nov,
       volume = {490},
       number = {1},
        pages = {368-376},
          doi = {10.1086/304869},
archivePrefix = {arXiv},
       eprint = {astro-ph/9706042},
 primaryClass = {astro-ph},
       adsurl = {https://ui.adsabs.harvard.edu/abs/1997ApJ...490..368C}
}

@INPROCEEDINGS{Drazkowska_2023,
       author = {{Dr{\k{a}}{\.z}kowska}, J. and {Bitsch}, B. and {Lambrechts}, M. and {Mulders}, G.~D. and {Harsono}, D. and {Vazan}, A. and {Liu}, B. and {Ormel}, C.~W. and {Kretke}, K. and {Morbidelli}, A.},
        title = "{Planet Formation Theory in the Era of ALMA and Kepler: from Pebbles to Exoplanets}",
    booktitle = {Protostars and Planets VII},
         year = 2023,
       editor = {{Inutsuka}, S. and {Aikawa}, Y. and {Muto}, T. and {Tomida}, K. and {Tamura}, M.},
       series = {Astronomical Society of the Pacific Conference Series},
       volume = {534},
        month = jul,
        pages = {717},
          doi = {10.48550/arXiv.2203.09759},
archivePrefix = {arXiv},
       eprint = {2203.09759},
 primaryClass = {astro-ph.EP},
       adsurl = {https://ui.adsabs.harvard.edu/abs/2023ASPC..534..717D}
}

@ARTICLE{Dutrey_2017,
       author = {{Dutrey}, A. and {Guilloteau}, S. and {Pi{\'e}tu}, V. and {Chapillon}, E. and {Wakelam}, V. and {Di Folco}, E. and {Stoecklin}, T. and {Denis-Alpizar}, O. and {Gorti}, U. and {Teague}, R. and et al.},
        title = "{The Flying Saucer: Tomography of the thermal and density gas structure of an edge-on protoplanetary disk}",
      journal = {\aap},
         year = 2017,
        month = nov,
       volume = {607},
          eid = {A130},
        pages = {A130},
          doi = {10.1051/0004-6361/201730645},
archivePrefix = {arXiv},
       eprint = {1706.02608},
 primaryClass = {astro-ph.SR},
       adsurl = {https://ui.adsabs.harvard.edu/abs/2017A&A...607A.130D}
}

@ARTICLE{Flores_2021,
       author = {{Flores}, C. and {Duch{\^e}ne}, G. and {Wolff}, S. and {Villenave}, M. and {Stapelfeldt}, K. and {Williams}, J.~P. and {Pinte}, C. and {Padgett}, D. and {Connelley}, M.~S. and {van der Plas}, G. and et al.},
        title = "{The Anatomy of an Unusual Edge-on Protoplanetary Disk. II. Gas Temperature and a Warm Outer Region}",
      journal = {\aj},
         year = 2021,
        month = may,
       volume = {161},
       number = {5},
          eid = {239},
        pages = {239},
          doi = {10.3847/1538-3881/abeb1e},
archivePrefix = {arXiv},
       eprint = {2103.02666},
 primaryClass = {astro-ph.SR},
       adsurl = {https://ui.adsabs.harvard.edu/abs/2021AJ....161..239F}
}

@ARTICLE{Swastik_2025,
       author = {{Swastik}, C. and {Wahhaj}, Z. and {Benisty}, M. and {Arora}, S. and {Ginski}, C. and {Ren}, B.~B. and {van Holstein}, R.~G. and {de Rosa}, R. and {Banyal}, R.~K. and {Tazaki}, R.},
        title = "{Imaging the LkCa 15 system in polarimetry and total intensity without self-subtraction artefacts}",
      journal = {\aap},
         year = 2026,
        month = feb,
       volume = {706},
          eid = {A312},
        pages = {A312},
          doi = {10.1051/0004-6361/202449743},
archivePrefix = {arXiv},
       eprint = {2512.18439},
 primaryClass = {astro-ph.EP},
       adsurl = {https://ui.adsabs.harvard.edu/abs/2026A&A...706A.312S}
}

@ARTICLE{Pohl_2017,
       author = {{Pohl}, A. and {Benisty}, M. and {Pinilla}, P. and {Ginski}, C. and {de Boer}, J. and {Avenhaus}, H. and {Henning}, Th. and {Zurlo}, A. and {Boccaletti}, A. and {Augereau}, J.-C. and {Birnstiel}, T. and {Dominik}, C. and {Facchini}, S. and {Fedele}, D. and {Janson}, M. and {Keppler}, M. and {Kral}, Q. and {Langlois}, M. and {Ligi}, R. and {Maire}, A.-L. and {M{\'e}nard}, F. and {Meyer}, M. and {Pinte}, C. and {Quanz}, S.~P. and {Sauvage}, J.-F. and {Sezestre}, {\'E}. and {Stolker}, T. and {Szul{\'a}gyi}, J. and {van Boekel}, R. and {van der Plas}, G. and {Villenave}, M. and {Baruffolo}, A. and {Baudoz}, P. and {Le Mignant}, D. and {Maurel}, D. and {Ramos}, J. and {Weber}, L.},
        title = "{The Circumstellar Disk HD 169142: Gas, Dust, and Planets Acting in Concert?}",
      journal = {\apj},
         year = 2017,
        month = nov,
       volume = {850},
       number = {1},
          eid = {52},
        pages = {52},
          doi = {10.3847/1538-4357/aa94c2},
archivePrefix = {arXiv},
       eprint = {1710.06485},
 primaryClass = {astro-ph.EP},
       adsurl = {https://ui.adsabs.harvard.edu/abs/2017ApJ...850...52P}
}

@ARTICLE{Andrews_2018,
       author = {{Andrews}, Sean M. and {Huang}, Jane and {P{\'e}rez}, Laura M. and {Isella}, Andrea and {Dullemond}, Cornelis P. and {Kurtovic}, Nicol{\'a}s T. and {Guzm{\'a}n}, Viviana V. and {Carpenter}, John M. and {Wilner}, David J. and {Zhang}, Shangjia and {Zhu}, Zhaohuan and {Birnstiel}, Tilman and {Bai}, Xue-Ning and {Benisty}, Myriam and {Hughes}, A. Meredith and {{\"O}berg}, Karin I. and {Ricci}, Luca},
        title = "{The Disk Substructures at High Angular Resolution Project (DSHARP). I. Motivation, Sample, Calibration, and Overview}",
      journal = {\apjl},
         year = 2018,
        month = dec,
       volume = {869},
       number = {2},
          eid = {L41},
        pages = {L41},
          doi = {10.3847/2041-8213/aaf741},
archivePrefix = {arXiv},
       eprint = {1812.04040},
 primaryClass = {astro-ph.SR},
       adsurl = {https://ui.adsabs.harvard.edu/abs/2018ApJ...869L..41A}
}

@ARTICLE{Aikawa_Nomura_2006,
       author = {{Aikawa}, Y. and {Nomura}, H.},
        title = "{Physical and Chemical Structure of Protoplanetary Disks with Grain Growth}",
      journal = {\apj},
         year = 2006,
        month = may,
       volume = {642},
       number = {2},
        pages = {1152-1162},
          doi = {10.1086/501114},
archivePrefix = {arXiv},
       eprint = {astro-ph/0601230},
 primaryClass = {astro-ph},
       adsurl = {https://ui.adsabs.harvard.edu/abs/2006ApJ...642.1152A}
}

@ARTICLE{Woitke_2016,
       author = {{Woitke}, P. and {Min}, M. and {Pinte}, C. and {Thi}, W.-F. and {Kamp}, I. and {Rab}, C. and {Anthonioz}, F. and {Antonellini}, S. and {Baldovin-Saavedra}, C. and {Carmona}, A. and {Dominik}, C. and {Dionatos}, O. and {Greaves}, J. and {G{\"u}del}, M. and {Ilee}, J.~D. and {Liebhart}, A. and {M{\'e}nard}, F. and {Rigon}, L. and {Waters}, L.~B.~F.~M. and {Aresu}, G. and {Meijerink}, R. and {Spaans}, M.},
        title = "{Consistent dust and gas models for protoplanetary disks. I. Disk shape, dust settling, opacities, and PAHs}",
      journal = {\aap},
         year = 2016,
        month = feb,
       volume = {586},
          eid = {A103},
        pages = {A103},
          doi = {10.1051/0004-6361/201526538},
archivePrefix = {arXiv},
       eprint = {1511.03431},
 primaryClass = {astro-ph.EP},
       adsurl = {https://ui.adsabs.harvard.edu/abs/2016A&A...586A.103W}
}

@ARTICLE{Vyasunin_2011,
       author = {{Vasyunin}, A.~I. and {Wiebe}, D.~S. and {Birnstiel}, T. and {Zhukovska}, S. and {Henning}, T. and {Dullemond}, C.~P.},
        title = "{Impact of Grain Evolution on the Chemical Structure of Protoplanetary Disks}",
      journal = {\apj},
         year = 2011,
        month = feb,
       volume = {727},
       number = {2},
          eid = {76},
        pages = {76},
          doi = {10.1088/0004-637X/727/2/76},
archivePrefix = {arXiv},
       eprint = {1011.4420},
 primaryClass = {astro-ph.GA},
       adsurl = {https://ui.adsabs.harvard.edu/abs/2011ApJ...727...76V}
}

@ARTICLE{Bruderer_2012,
       author = {{Bruderer}, S. and {van Dishoeck}, E.~F. and {Doty}, S.~D. and {Herczeg}, G.~J.},
        title = "{The warm gas atmosphere of the HD 100546 disk seen by Herschel. Evidence of a gas-rich, carbon-poor atmosphere?}",
      journal = {\aap},
         year = 2012,
        month = may,
       volume = {541},
          eid = {A91},
        pages = {A91},
          doi = {10.1051/0004-6361/201118218},
archivePrefix = {arXiv},
       eprint = {1201.4860},
 primaryClass = {astro-ph.SR},
       adsurl = {https://ui.adsabs.harvard.edu/abs/2012A&A...541A..91B}
}

@ARTICLE{Bruderer_2013,
       author = {{Bruderer}, Simon},
        title = "{Survival of molecular gas in cavities of transition disks. I. CO}",
      journal = {\aap},
         year = 2013,
        month = nov,
       volume = {559},
          eid = {A46},
        pages = {A46},
          doi = {10.1051/0004-6361/201321171},
archivePrefix = {arXiv},
       eprint = {1308.2966},
 primaryClass = {astro-ph.SR},
       adsurl = {https://ui.adsabs.harvard.edu/abs/2013A&A...559A..46B}
}

@ARTICLE{Youdin_Lithwick_2007,
       author = {{Youdin}, Andrew N. and {Lithwick}, Yoram},
        title = "{Particle stirring in turbulent gas disks: Including orbital oscillations}",
      journal = {\icarus},
         year = 2007,
        month = dec,
       volume = {192},
       number = {2},
        pages = {588-604},
          doi = {10.1016/j.icarus.2007.07.012},
archivePrefix = {arXiv},
       eprint = {0707.2975},
 primaryClass = {astro-ph},
       adsurl = {https://ui.adsabs.harvard.edu/abs/2007Icar..192..588Y}
}

@ARTICLE{Rosotti_2025,
       author = {{Rosotti}, Giovanni P. and {Longarini}, Cristiano and {Paneque-Carre{\~n}o}, Teresa and {Cataldi}, Gianni and {Galloway-Sprietsma}, Maria and {Andrews}, Sean M. and {Bae}, Jaehan and {Barraza-Alfaro}, Marcelo and {Benisty}, Myriam and {Curone}, Pietro and {Czekala}, Ian and {Facchini}, Stefano and {Fasano}, Daniele and {Flock}, Mario and {Fukagawa}, Misato and {Garg}, Himanshi and {Hall}, Cassandra and {Huang}, Jane and {Ilee}, John D. and {Izquierdo}, Andr{\'e}s F. and {Kanagawa}, Kazuhiro and {Lesur}, Geoffroy and {Lodato}, Giuseppe and {Loomis}, Ryan A. and {Orihara}, Ryuta and {Pinte}, Christophe and {Price}, Daniel J. and {Stadler}, Jochen and {Teague}, Richard and {Fernandez}, Gaylor Wafflard- and {Winter}, Andrew J. and {W{\"o}lfer}, Lisa and {Yen}, Hsi-Wei and {Yoshida}, Tomohiro C. and {Zawadzki}, Brianna},
        title = "{exoALMA. XV. Interpreting the Height of CO Emission Layer}",
      journal = {\apjl},
         year = 2025,
        month = may,
       volume = {984},
       number = {1},
          eid = {L20},
        pages = {L20},
          doi = {10.3847/2041-8213/adc42e},
archivePrefix = {arXiv},
       eprint = {2504.20012},
 primaryClass = {astro-ph.EP},
       adsurl = {https://ui.adsabs.harvard.edu/abs/2025ApJ...984L..20R}
}

@ARTICLE{Beckwith_1990,
       author = {{Beckwith}, Steven V.~W. and {Sargent}, Anneila I. and {Chini}, Rolf S. and {Guesten}, Rolf},
        title = "{A Survey for Circumstellar Disks around Young Stellar Objects}",
      journal = {\aj},
         year = 1990,
        month = mar,
       volume = {99},
        pages = {924},
          doi = {10.1086/115385},
       adsurl = {https://ui.adsabs.harvard.edu/abs/1990AJ.....99..924B}
}

@ARTICLE{Teague_2025,
       author = {{Teague}, Richard and {Benisty}, Myriam and {Facchini}, Stefano and {Fukagawa}, Misato and {Pinte}, Christophe and {Andrews}, Sean M. and {Bae}, Jaehan and {Barraza-Alfaro}, Marcelo and {Cataldi}, Gianni and {Cuello}, Nicol{\'a}s and {Curone}, Pietro and {Czekala}, Ian and {Fasano}, Daniele and {Flock}, Mario and {Galloway-Sprietsma}, Maria and {Garg}, Himanshi and {Hall}, Cassandra and {Hammond}, Iain and {Hilder}, Thomas and {Huang}, Jane and {Ilee}, John D. and {Izquierdo}, Andr{\'e}s F. and {Kanagawa}, Kazuhiro and {Lesur}, Geoffroy and {Lodato}, Giuseppe and {Longarini}, Cristiano and {Loomis}, Ryan A. and {Masset}, Fr{\'e}d{\'e}ric and {Menard}, Francois and {Orihara}, Ryuta and {Price}, Daniel J. and {Rosotti}, Giovanni and {Stadler}, Jochen and {Testi}, Leonardo and {Yen}, Hsi-Wei and {Wafflard-Fernandez}, Gaylor and {Wilner}, David J. and {Winter}, Andrew J. and {W{\"o}lfer}, Lisa and {Yoshida}, Tomohiro C. and {Zawadzki}, Brianna},
        title = "{exoALMA. I. Science Goals, Project Design, and Data Products}",
      journal = {\apjl},
         year = 2025,
        month = may,
       volume = {984},
       number = {1},
          eid = {L6},
        pages = {L6},
          doi = {10.3847/2041-8213/adc43b},
archivePrefix = {arXiv},
       eprint = {2504.18688},
 primaryClass = {astro-ph.EP},
       adsurl = {https://ui.adsabs.harvard.edu/abs/2025ApJ...984L...6T}
}

@ARTICLE{Byrne_2026,
       author = {{Byrne}, J. and {Ginski}, C. and {van Capelleveen}, R.~F. and {Fitzgerald}, N. and {Garufi}, A. and {Coyne}, C. and {Lawlor}, C. and {McLachlan}, D.},
        title = "{Vertical structure of protoplanetary disks in scattered light: A large-sample analysis}",
      journal = {\aap},
         year = 2026,
        month = may,
       volume = {710},
          eid = {A38},
        pages = {A38},
          doi = {10.1051/0004-6361/202557880},
archivePrefix = {arXiv},
       eprint = {2603.05599},
 primaryClass = {astro-ph.EP},
       adsurl = {https://ui.adsabs.harvard.edu/abs/2026A&A...710A..38B}
}

@ARTICLE{Dalessio_1998,
       author = {{D'Alessio}, Paola and {Cant{\"o}}, Jorge and {Calvet}, Nuria and {Lizano}, Susana},
        title = "{Accretion Disks around Young Objects. I. The Detailed Vertical Structure}",
      journal = {\apj},
         year = 1998,
        month = jun,
       volume = {500},
       number = {1},
        pages = {411-427},
          doi = {10.1086/305702},
archivePrefix = {arXiv},
       eprint = {astro-ph/9806060},
 primaryClass = {astro-ph},
       adsurl = {https://ui.adsabs.harvard.edu/abs/1998ApJ...500..411D}
}

@ARTICLE{Garufi_2026,
       author = {{Garufi}, A. and {Ginski}, C. and {Benisty}, M. and {Vioque}, M. and {Winter}, A. and {Huang}, J. and {Manara}, C.~F. and {Dominik}, C.},
        title = "{Planet-forming disks and their environment across regions and time from the full NIR census}",
      journal = {\aap},
         year = 2026,
        month = may,
       volume = {709},
          eid = {A269},
        pages = {A269},
          doi = {10.1051/0004-6361/202558522},
archivePrefix = {arXiv},
       eprint = {2603.01703},
 primaryClass = {astro-ph.SR},
       adsurl = {https://ui.adsabs.harvard.edu/abs/2026A&A...709A.269G}
}

@ARTICLE{Pezzotta_2026,
       author = {{Pezzotta}, V. and {Facchini}, S. and {Izquierdo}, A.~F. and {Lodato}, G. and {Longarini}, C. and {Bae}, J. and {Galloway-Sprietsma}, M. and {Pinte}, C. and {Law}, C.~J. and {Paneque-Carre{\~n}o}, T.},
        title = "{Extending dynamical mass measurements: probing GI as a possible origin of mm-dust spirals}",
      journal = {arXiv e-prints},
         year = 2026,
        month = jul,
          eid = {arXiv:2607.15923},
        pages = {arXiv:2607.15923},
archivePrefix = {arXiv},
       eprint = {2607.15923},
 primaryClass = {astro-ph.EP},
       adsurl = {https://ui.adsabs.harvard.edu/abs/2026arXiv260715923P}
}

\begin{appendix}

\section{Nonisotropic scattering}
\label{sec: appendix-non-isotropic}
\begin{figure}
    \centering
    \includegraphics[width=\hsize]{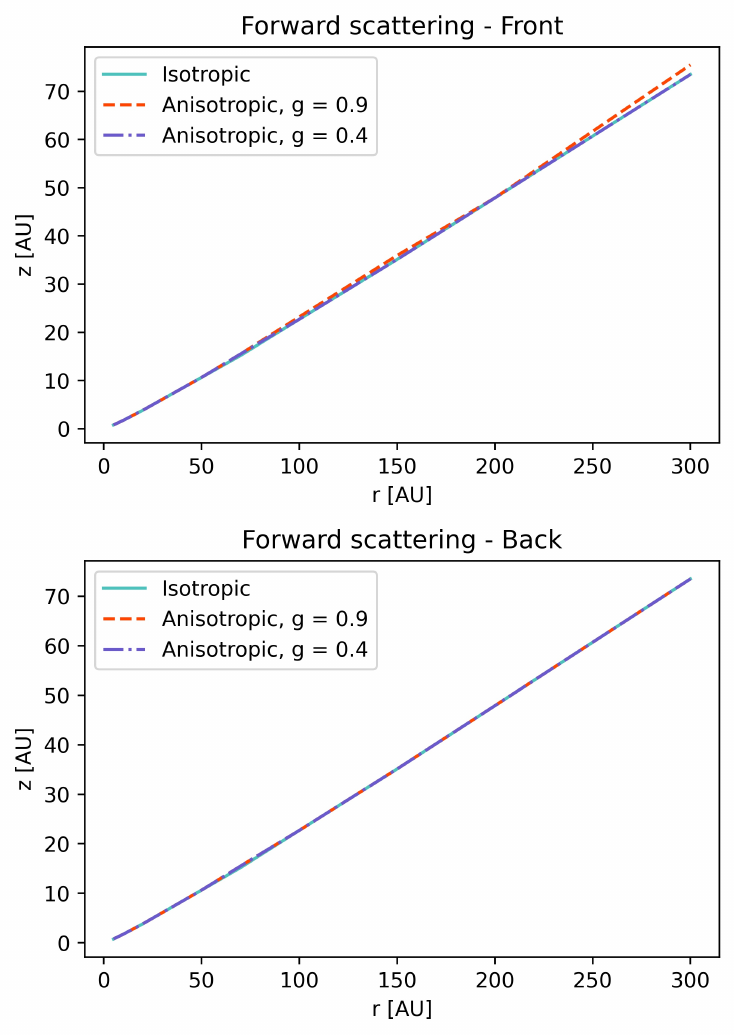}
    \caption{Scattering height considering forward scattering and isotropic scattering.}
    \label{fig:non-isotropic scattering}
\end{figure}
Here, we test the isotropic scattering approximation by verifying whether considering nonisotropic scattering, with various degree of anisotropy, significantly affects the location of the scattering surface. Scattering anisotropy is described by the scattering phase function, i.e. the probability of scattering into the direction $\mu=\cos\psi$, where $\psi$ is the deflection with respect to the photon incoming direction. If the photon is scattered backward $\psi = \pi$, if forward $\psi = 0$.

We use the Henyen-Greenstein parametrization of the phase function \citep{Henyey_Greenstein_1941} via
\begin{equation}
    \beta(g,\psi)=\frac{1}{4\pi}\frac{1-g^2}{(1+g^2-2g\cos\psi)^{3/2}}
,\end{equation}
where $g \in [-1,1]$ parametrizes the degree of anisotropy. Considering the scattering of the stellar light with the protoplanetary disk, the maximum scattering angle corresponds to the near side of the disk, and the maximum angle to the far side. Therefore, we expect to observe the maximum difference in the height between the far side and the near side of the disk. Consequently, assuming an anisotropy factor g intrinsic of the dust, we compare the scattering height between the far side and the near side and see if there is a significant difference with respect to the isotropic scattering case.

The contribution of the phase function can be inserted into Eq. \ref{eq: Paola_Dalessio}, which becomes
\begin{equation}
    \frac{dI_{\nu}}{ds}=\sigma_{\nu}\beta(g,\psi)B_{\nu}(T_*)W(r)e^{-\tau_{\nu,*}}e^{-\tau_{\nu}(s)}.
\end{equation}
We compute the scattering surface for various degrees of anisotropy $g$, both in the case of forward and backward scattering. Figure \ref{fig:non-isotropic scattering} shows an example in the case of forward scattering for $i = 20^{\circ}$. The scattering surface is essentially unaffected by the phase function, even with high degrees of anisotropy. The reason behind this irrelevance is that $\tau_1$ is such a steeply varying function of $z$ (Fig. \ref{fig:tau_12_sigma}) that the phase function does not move the $\tau_1\approx 1$ surface by much. In general, the scattering surface depends on azimuth, but the difference between the two sides, and the difference between the non isotropic and isotropic scattering is at most $2$ AU, so it is negligible, compared to the typical errors in measuring the scattering height. In the case of higher inclinations this effect can become more important, however we already discussed that our model breaks down at high inclinations.
It is therefore reasonable to ignore the scattering anisotropy. 

\section{Isothermal approximation}
\label{sec:vert_temp_gradient}
Here we test whether it would be reasonable to assume that the disk is vertically isothermal. 
\begin{figure}[h!]
    \centering
    \includegraphics[width=\hsize]{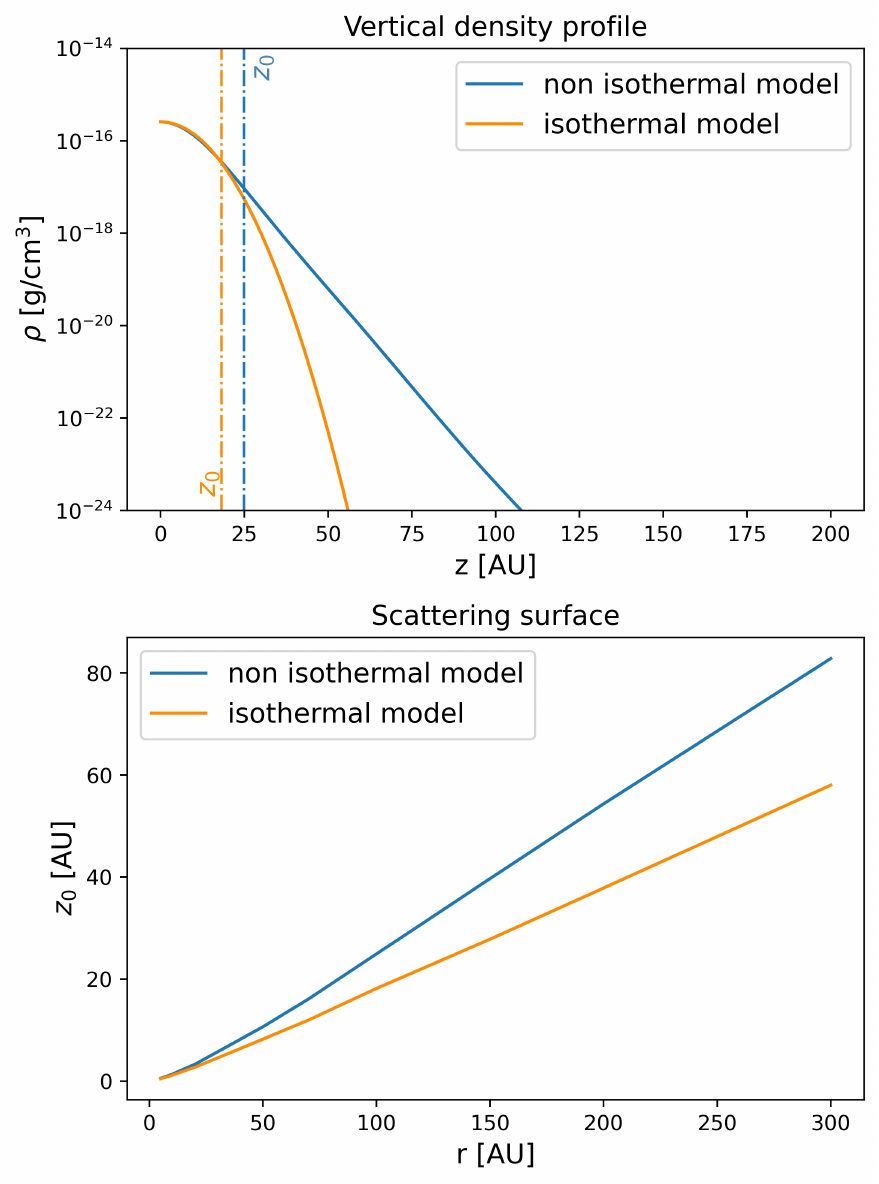}
    \caption{Comparison between the isothermal and nonisothermal case. \textit{Top}: Vertical density profile at $r = 100$ AU. \textit{Bottom}: Scattering surface.}
    \label{fig:isotermo_vs_non}
\end{figure}
We model the disk as isothermal: the scale height can be computed from the midplane temperature as in Eq \ref{eq: H_scale}, and the density profile becomes 
\begin{equation}
    \rho(z)=\rho_0\exp\left[-\frac{z^2}{2H^2(r)}\right].
\end{equation}
Figure \ref{fig:isotermo_vs_non} shows the comparison between the stratified temperature case and the isothermal approximation for LkCa15: in the top panel we show the vertical density profile at 100 AU, and we point out that in the non isothermal case the disk is more flared. Consequently the scattering surface is higher, as shown in the second panel of Fig. \ref{fig:isotermo_vs_non}. The difference accounts for more than 20 AU, so the isothermal approximation while convenient is not adequate. Practically speaking, this means that when modeling the observations described in Sect. \ref{sec: results}, we would require observational measurements of the temperature structure.

\section{Opacity}
\label{appendix-opacity}
When inverting the model, there is a clear degeneracy between dust mass and opacity. In our work, we used the \citet{Ossenkopf_Henning_1994} opacity model. 
\begin{figure}
    \centering
    \includegraphics[width=\hsize]{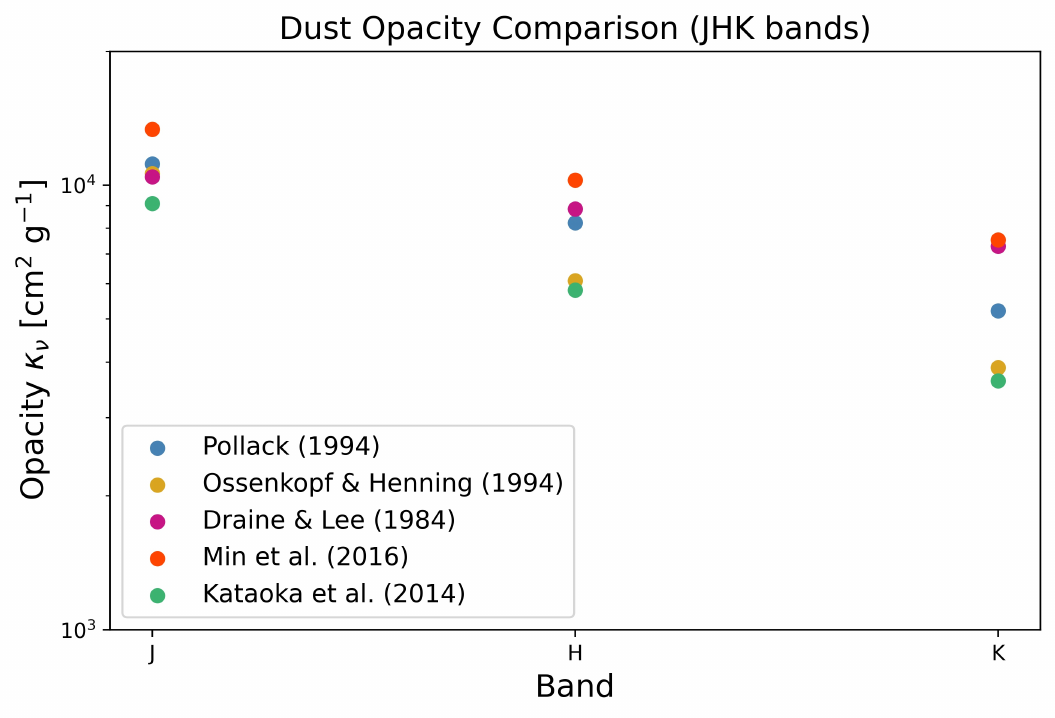}
    \caption{Opacity for different dust models.}
    \label{fig:opacity_compositions}
\end{figure}
Figure~\ref{fig:opacity_compositions} shows the dust opacity in the J ($\lambda \approx 1.2,\mu$m), H ($\lambda \approx 1.6,\mu$m), and K ($\lambda \approx 2.2,\mu$m) bands, computed with \texttt{Optool} \citep{Dominik_2021}, considering grains up to 1 $\mu$m. The comparison includes five widely adopted opacity models with different compositions and porosities \citep{Draine_Lee_1984, Ossenkopf_Henning_1994, Pollack_1994, Kataoka_2015, Min_2016}. 

This comparison highlights the significant uncertainties associated with dust opacities, which can vary between models. As a result, it is generally not meaningful to directly compare the scattering heights derived from observations of different disks taken in different bands, since the variations in opacity can dominate over wavelength-dependent effects. However, if multi-band scattered-light observations are available for the same disk, differences in the derived scattering height across bands can provide valuable constraints on the grain size distribution.
\section{Grain size distribution plots}
Here, we present the constraints on the grain size distribution for the rest of the disks.
\label{appendix-gsd}
\begin{figure}[h!]
    \centering
    \includegraphics[width=\hsize]{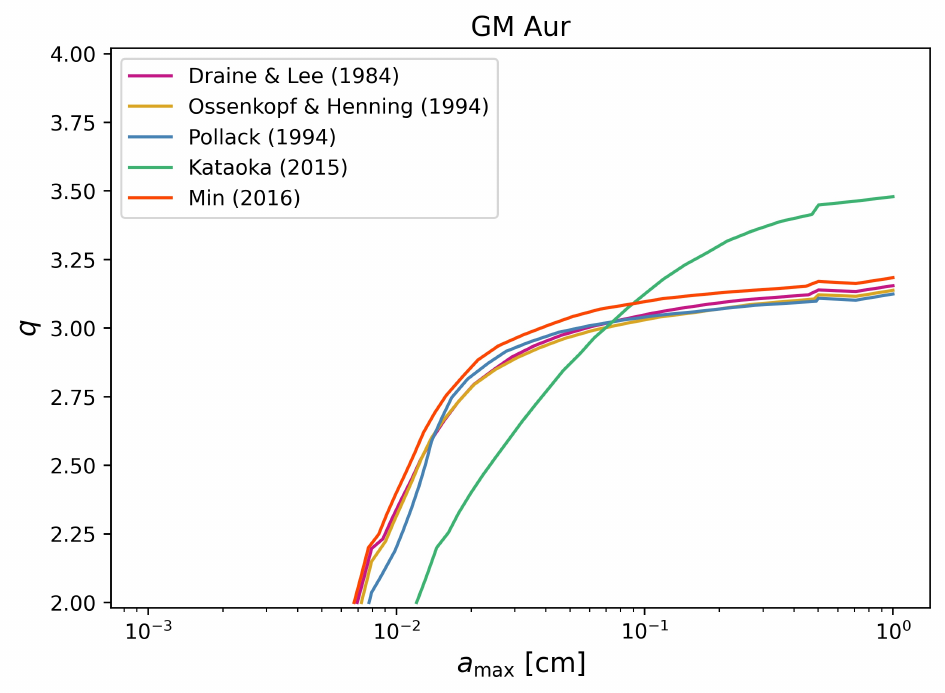}
    \caption{$f=1$ curve as computed in Sect. \ref{sec: gsd} for GM Aur.}
    \label{fig:GMAur}
\end{figure}
\begin{figure}[h!]
    \centering
    \includegraphics[width=\hsize]{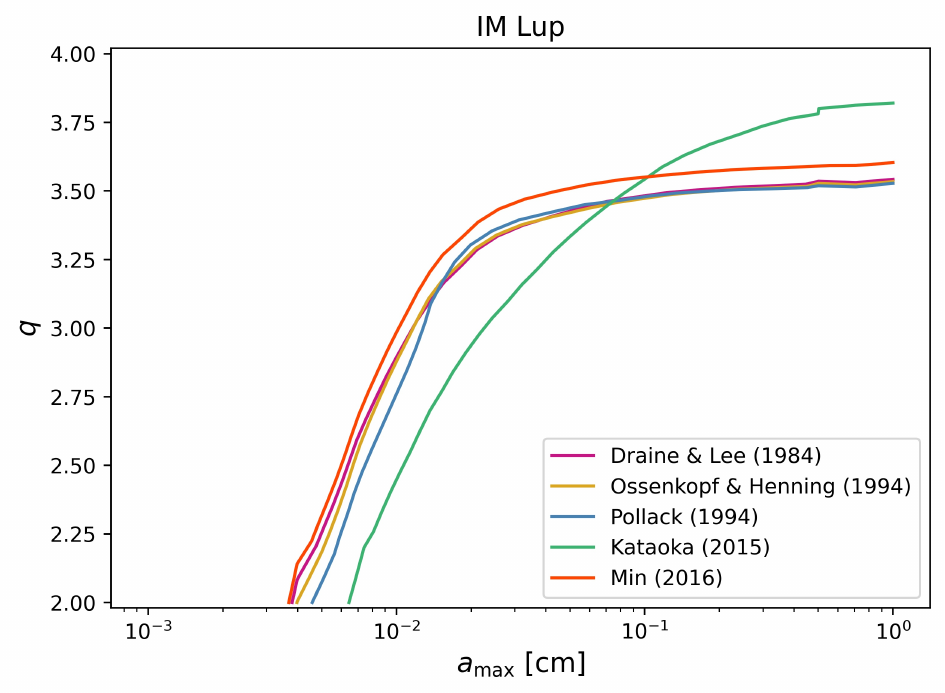}
    \caption{$f=1$ curve as computed in Sect. \ref{sec: gsd} for IM Lup.}
    \label{fig:IMLup}
\end{figure}
\begin{figure}[h!]
    \centering
    \includegraphics[width=\hsize]{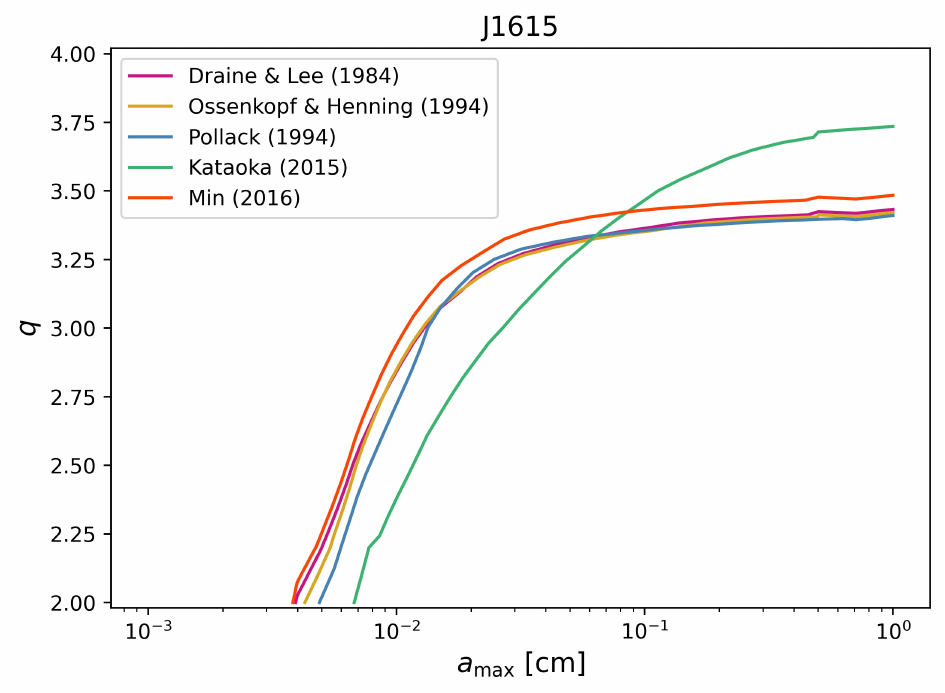}
    \caption{$f=1$ curve as computed in Sect. \ref{sec: gsd} for J1615.}
    \label{fig:J1615}
\end{figure}
\begin{figure}[h!]
    \centering
    \includegraphics[width=\hsize]{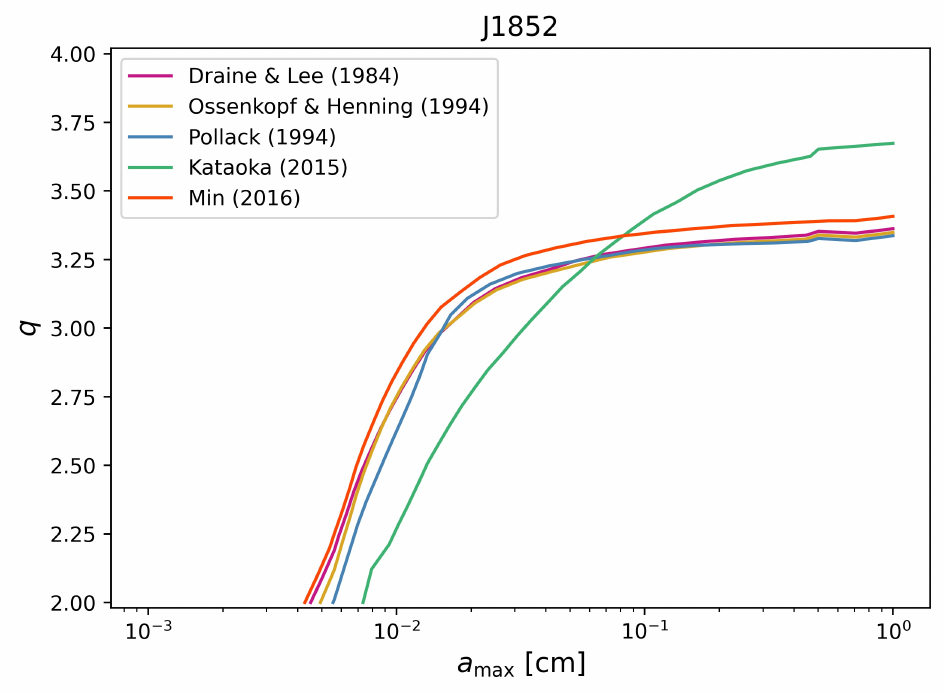}
    \caption{$f=1$ curve as computed in Sect. \ref{sec: gsd} for J1852.}
    \label{fig:J1852}
\end{figure}
\begin{figure}[h!]
    \centering
    \includegraphics[width=\hsize]{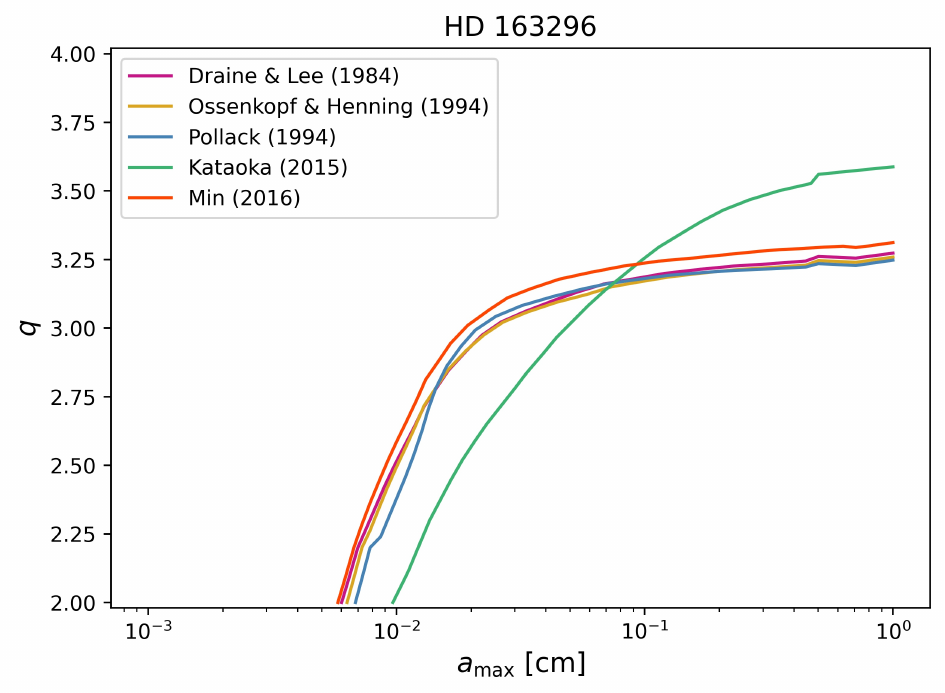}
    \caption{$f=1$ curve as computed in Sect. \ref{sec: gsd} for HD 163296.}
    \label{fig:GMAur}
\end{figure}
\begin{figure}[h!]
    \centering
    \includegraphics[width=\hsize]{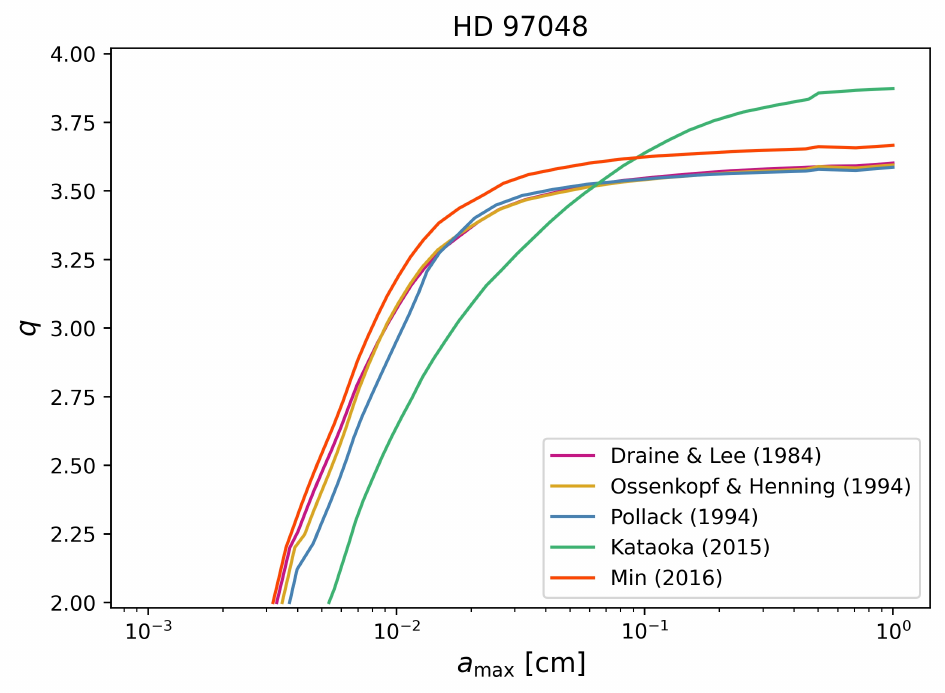}
    \caption{$f=1$ curve as computed in Sect. \ref{sec: gsd} for HD 97048.}
    \label{fig:HD97048}
\end{figure}
\begin{figure}[h!]
    \centering
    \includegraphics[width=\hsize]{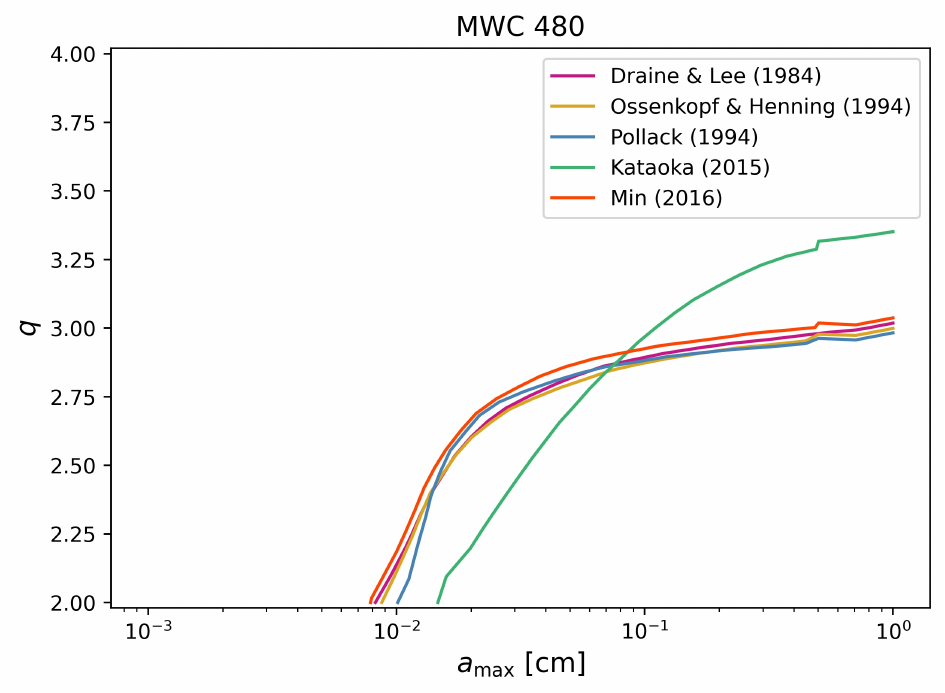}
    \caption{$f=1$ curve as computed in Sect. \ref{sec: gsd} for MWC 480.}
    \label{fig:MWC480}
\end{figure}
\begin{figure}[h!]
    \centering
    \includegraphics[width=\hsize]{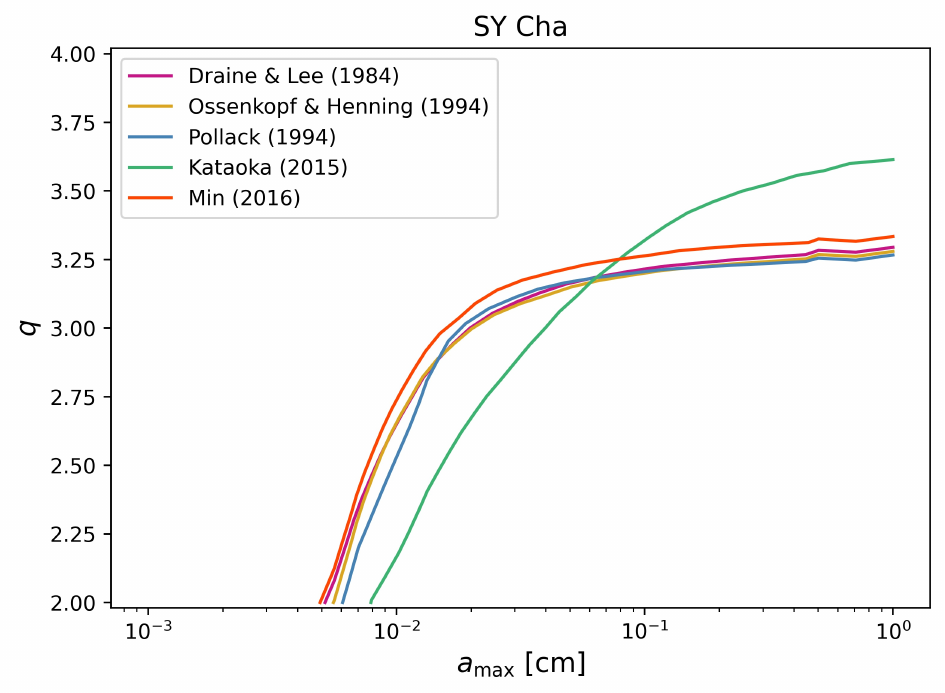}
    \caption{$f=1$ curve as computed in Sect. \ref{sec: gsd} for SY Cha.}
    \label{fig:SYCHA}
\end{figure}
\begin{figure}[h!]
    \centering
    \includegraphics[width=\hsize]{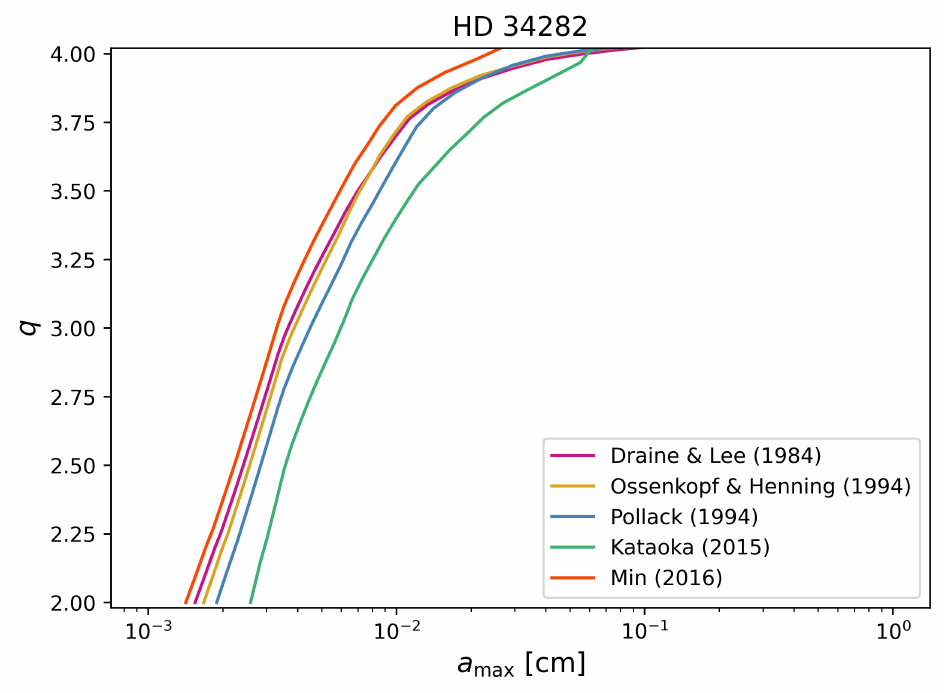}
    \caption{$f=1$ curve as computed in Sect. \ref{sec: gsd} for HD 34282.}
    \label{fig:HD34282}
\end{figure}

\section{Surface density profile}
\label{appendix: surface density}
In Eq. \ref{eq: inverse}, we assume that the surface density profile scales as $r^{-1}$ before the exponential cutoff. Here, we quantify the impact of this assumption. Considering a more general \citet{LyndenBell_Pringle_1974} profile,
\begin{equation}
    \Sigma(r)=\Sigma_0\left(\frac{r}{rc}\right)^{-\gamma}\exp\left[-\left(\frac{r}{r_c}\right)^{2-\gamma}\right],
\end{equation}
we can compute the small dust mass varying the exponent $\gamma$ between 0.5 and 1.5. The mass difference amounts to a factor of $\approx 2.5$ and here we report the small dust mass measurements for $\gamma=0.5$ and $\gamma=1.5$.
\begin{table}[h!]
\centering
\caption{Small dust mass measurements corrected for different $\gamma$ values. }
\begin{tabular}{cccc}
\hline
\hline
Disk & Mass $\gamma=0.5$ & Mass $\gamma=1.0$ & Mass $\gamma=1.5$ \\
 &  $\log(M_{sd}/M_{\oplus})$ & $\log(M_{sd}/M_{\oplus})$ & $\log(M_{sd}/M_{\oplus})$ \\
\hline
V4046 Sgr ring 1 & $-2.24 \pm 0.19$ & $-2.30 \pm 0.20$ & $-2.47 \pm 0.22$ \\
V4046 Sgr ring 2  & $-1.22 \pm 0.10$ & $-1.25 \pm 0.10$ & $-1.35 \pm 0.11$ \\
RXJ 1615 ring 1  & $-1.37 \pm 0.29$ & $-1.41 \pm 0.30$ & $-1.52 \pm 0.32$ \\
RXJ 1615 ring 2 & $-0.89 \pm 0.15$ & $-0.91 \pm 0.15$ & $-0.98 \pm 0.16$ \\
RXJ 1615 ring 3 & $-0.62 \pm 0.04$ & $-0.64 \pm 0.04$ & $-0.69 \pm 0.04$ \\
IM Lup ring 1  & $-0.73 \pm 0.40$ & $-0.75 \pm 0.41$ & $-0.81 \pm 0.44$ \\
IM Lup ring 2& $-0.68 \pm 0.45$ & $-0.70 \pm 0.46$ & $-0.75 \pm 0.49$ \\
IM Lup ring 3 & $-0.17 \pm 0.32$ & $-0.17 \pm 0.33$ & $-0.18 \pm 0.35$ \\
IM Lup ring 4  & $0.01 \pm 0.40$ & $0.01 \pm 0.41$ & $0.01 \pm 0.44$ \\
HD 163296 & $-0.91 \pm 0.30$ & $-0.93 \pm 0.31$ & $-1.00 \pm 0.33$ \\
LkCa 15 & $-0.71 \pm 0.31$ & $-0.73 \pm 0.32$ & $-0.79 \pm 0.34$ \\
MWC 480 & $-1.42 \pm 0.54$ & $-1.46 \pm 0.55$ & $-1.57 \pm 0.59$ \\
J1852 &  $-1.30 \pm 0.87$ & $-1.33 \pm 0.89$ & $-1.43 \pm 0.96$ \\
SY Cha & $-1.32 \pm 1.90$ & $-1.36 \pm 1.95$ & $-1.46 \pm 2.10$ \\
HD 97048 ring 1 & $0.16 \pm 0.61$ & $0.16 \pm 0.63$ & $0.17 \pm 0.68$ \\
HD 97048 ring 2 & $0.67 \pm 0.60$ & $0.69 \pm 0.62$ & $0.74 \pm 0.67$ \\
HD 97048 ring 3 & $0.40 \pm 0.59$ & $0.41 \pm 0.61$ & $0.44 \pm 0.66$ \\
GM Aur &  $-1.41 \pm 0.12$ & $-1.45 \pm 0.12$ & $-1.56 \pm 0.13$ \\
\hline
\hline
\end{tabular}
\tablefoot{The measurement refers to the cumulative mass at the radius of the given ring.}
\label{tab:dust_mass_gamma}
\end{table}

\end{appendix}

\end{document}